\documentclass[twocolumn]{aastex631}
\usepackage{amsmath}
\usepackage{graphicx}
\usepackage[utf8]{inputenc}
\usepackage{makecell}
\usepackage{chngcntr}

\usepackage{placeins}
\usepackage{multirow}
\usepackage{enumitem}
\submitjournal{AAS Journal}

\shorttitle{Synergy of CHES and HWO for detecting habitable planets.}
\shortauthors{Chunhui Bao et al.}

\graphicspath{{./}{figs/}}

\begin{document}

\title{Closeby Habitable Exoplanet Survey (CHES). \uppercase\expandafter{\romannumeral4}. Synergy between astrometry and direct imaging missions of the Habitable World Observatory for detecting Earth-like planets}

\author[0009-0001-6688-8984]{Chunhui Bao}
\affiliation{CAS Key Laboratory of Planetary Sciences, Purple Mountain Observatory, Chinese Academy of Sciences, Nanjing 210023, China;jijh@pmo.ac.cn}
\affiliation{School of Astronomy and Space Science, University of Science and Technology of China, Hefei 230026, China}

\author[0000-0002-9260-1537]{Jianghui Ji}
\affiliation{CAS Key Laboratory of Planetary Sciences, Purple Mountain Observatory, Chinese Academy of Sciences, Nanjing 210023, China;jijh@pmo.ac.cn}
\affiliation{School of Astronomy and Space Science, University of Science and Technology of China, Hefei 230026, China}
\affiliation{CAS Center for Excellence in Comparative Planetology, Hefei 230026, China}

\author[0000-0002-0785-5015]{Dongjie Tan}
\affiliation{CAS Key Laboratory of Planetary Sciences, Purple Mountain Observatory, Chinese Academy of Sciences, Nanjing 210023, China;jijh@pmo.ac.cn}
\affiliation{School of Astronomy and Space Science, University of Science and Technology of China, Hefei 230026, China}

\author[0000-0003-0740-5433]{Guo Chen}
\affiliation{CAS Key Laboratory of Planetary Sciences, Purple Mountain Observatory, Chinese Academy of Sciences, Nanjing 210023, China;jijh@pmo.ac.cn}
\affiliation{School of Astronomy and Space Science, University of Science and Technology of China, Hefei 230026, China}

\author[0009-0006-2563-2479]{Xiumin Huang}
\affiliation{CAS Key Laboratory of Planetary Sciences, Purple Mountain Observatory, Chinese Academy of Sciences, Nanjing 210023, China;jijh@pmo.ac.cn}
\affiliation{School of Astronomy and Space Science, University of Science and Technology of China, Hefei 230026, China}

\author[0000-0002-4859-259X]{Su Wang}
\affiliation{CAS Key Laboratory of Planetary Sciences, Purple Mountain Observatory, Chinese Academy of Sciences, Nanjing 210023, China;jijh@pmo.ac.cn}
\affiliation{CAS Center for Excellence in Comparative Planetology, Hefei 230026, China}

\author[0000-0001-8162-3485]{Yao Dong}
\affiliation{CAS Key Laboratory of Planetary Sciences, Purple Mountain Observatory, Chinese Academy of Sciences, Nanjing 210023, China;jijh@pmo.ac.cn}
\affiliation{CAS Center for Excellence in Comparative Planetology, Hefei 230026, China}

\begin{abstract}
The detection and characterization of habitable planets around nearby stars persist as one of the foremost objectives in contemporary astrophysics. This work investigates the synergistic integration of astrometric and direct imaging techniques by capitalizing on the complementary capabilities of the Closeby Habitable Exoplanet Survey (CHES) and Habitable Worlds Observatory (HWO). Planetary brightness and position vary over time due to phase effects and orbital architectures, information that can be precisely provided by CHES's astrometric measurements. By combining the precise orbital constraints from CHES with the imaging capabilities of HWO, we evaluate the improvements in detection efficiency, signal-to-noise ratio and overall planet yield. Completeness is quantified as the fraction of injected planets that are successfully detected, while yields are estimated for various scenarios using terrestrial planet occurrence rates derived from the Kepler dataset. Our results indicate that prior astrometric data significantly enhance detection efficiency. Under the adopted detection limit, our analysis indicates that prior CHES observations can increase completeness by approximately 10\% and improve detection efficiency by factors ranging from two to thirty. The findings underscore the importance of interdisciplinary approaches in the search for and characterization of habitable worlds.

\end{abstract}


\section{Introduction} \label{sec:intro}

The detection of exoplanets has advanced rapidly in recent years, uncovering a diverse range of planetary systems that challenge our understanding of planetary formation and evolution. Among the various methods used for exoplanet discovery, direct imaging and astrometry provide unique advantages, particularly for detecting and characterizing planets in wider orbits. Direct imaging enables the spatial separation of a planet from its host star, facilitating studies of planetary atmospheres, climates, and surface compositions through spectral analysis \citep{Seager2010}. Modern telescopes or facilities, such as the Very Large Telescope (VLT) \citep{Quanz2015}, the James Webb Space Telescope (JWST) \citep{Calissendorff2023}, and the Gemini Planet Imager (GPI) \citep{Macintosh2014}, have discovered nearly 100 planets through direct imaging. The atmospheric composition can be determined by analyzing thermal emission spectra. However, limitations in starlight suppression techniques pose significant challenges to detect habitable rocky planets, which require contrast $<10^{-10}$ in the optical band and angular separation on the order of a hundred of milli-arcseconds. Currently, most planets detected through imaging are self-luminous giants observed via thermal emission at near- and mid-infrared wavelengths, with their orbital parameters and masses often poorly constrained.

Motivated by advancements in coronagraph or starshade technology, several next-generation direct imaging missions have been proposed, with the detection of habitable planets as a primary objective. These missions include the {Habitable World Observatory (HWO) \citep{Mamajek2024}}, the {Nancy Grace Roman Space Telescope (formerly known as WFIRST) \citep{Poberezhskiy2020}}, the {Large Interferometer For Exoplanets (LIFE) \citep{Quanz2022, Konrad2022}}, the {Miyin \citep{jianghui2020china, Wang2024}}, and the Cool Planet Imaging Coronagraph (CPI-C), planned for deployment on the China Space Station Telescope (CSST) \citep{CSST2015, Feng2024}.

Another key challenge is that a planet's detectability varies over time due to its orbital elements and atmospheric properties \citep{Greco2015}, requiring multiple observations to achieve high completeness in direct imaging \citep{Stark2014, Stark2024}. Astrometry, on the other hand, provides precise measurements of host star positions over time, enabling the retrieval of key planetary orbital parameters, such as inclination, eccentricity, and period, which are essential for understanding planetary system architecture. When combined with direct imaging, astrometry enhances the robustness of exoplanet detections and reduce uncertainties in planetary characteristics, offering a more comprehensive view of each system. For instance, the Gaia satellite has already discovered dozens of giant planets \citep{Gaia2023}, and is expected to detect tens of thousands of during a 10-year mission \citep{Perryman2014}. However, Gaia's precision in astrometry limits its ability to detect Earth-like planets. To address this gap, several upcoming missions aim to search for terrestrial planets using astrometry with micro-arcsecond precision. These include the Theia mission \citep{TheiaCollaboration2017}, the Closeby Habitable Exoplanet Survey (CHES) mission \citep{Ji2022, Ji2024}, and the Small-JASMINE mission \citep{Kawata2023}. Particularly, the CHES mission is designed to observe nearly a hundred nearby solar-type stars, many of which are also the targets of HWO, as both are more sensitive to the nearby stars.

Given the unique strengths of direct imaging and astrometry, their combination has been extensively studied. Several studies have shown that multi-epoch observations from direct imaging can aid in retrieving orbital parameters through astrometry \citep{Guimond2019, Bruna2023}. Conversely, astrometry or radial velocity measurements can constrain planetary orbits, improving the efficiency and accuracy of imaging \citep{Salvador2024}. Furthermore, orbital architectures derived from astrometric data offer valuable insights into imaging missions. \citet{Kane2018} discuss the optimal observation epochs based on maximum angular separation determined from planetary orbital parameters. Furthermore, additional insights into planetary system stability and habitability from astrometry would enhance target selection for direct imaging missions \citep{Kane2024}. \citet{Davidson2011} explored the role of prior knowledge in optimizing observation scheduling. \citet{Savransky2009} and \citet{Shao2010} show that prior knowledge of planetary orbits can significantly enhance the performance of direct detection missions, leading to increased yields and reduced observation time. Similarly, \citet{Morgan2021} find that exoplanet yields can increase by approximately 30\% when incorporating prior information from radial velocity, allowing the same yield to be achieved in half the observing time. \citet{Plavchan2024} consider an idealized scenario in which observational timing is optimized with perfect prior knowledge of planetary ephemerides. Their theoretical derivations suggest that precursor knowledge does not significantly impact yields. However, prior observations can reduce on-sky time by a factor of five to ten for a single visit observation.

This work investigates the synergy between direct imaging and astrometry in planetary detection, focusing on improvements in detection efficiency and accuracy, particularly terrestrial planets in habitable zones. CHES and HWO serve as prototypes for astrometry and imaging missions, respectively, due to their overlapping passbands and common targets. In this study, we assume that CHES observes the targets in advance, providing critical planetary information to guide HWO's follow-up observations. We adopt a practical model to simulate prior observations by CHES and use this information to optimize observational strategies for multi-planet systems, accounting for observational errors. Additionally, since direct imaging sensitivity depends on planetary radius and orbital distance, we divide the parameter space into multiple cells. Using Monte Carlo simulations, we estimate the detection completeness of terrestrial planets with varying radii and semi-major axes across all HWO targets. We find that the completeness can be increased by about 10\% for most targets with prior knowledge. Moreover, incorporating prior astrometric information from CHES leads to a slight increase in the predicted yield of habitable planets. In particular, the number of detected planets increases by five to ten compared to scenarios lacking prior astrometric observations, {yielding approximately 37 and 47 planets located within the conservative and optimistic habitable zones, respectively, based on the low-occurrence model \citep{Kopparapu2014,Bryson2021}}. Based on our results, we establish priority rankings for all target stars, offering a valuable reference for optimizing target selection in future missions.

The remainder of this work is organized as follows. Section \ref{sect:Mission} describes the observation modes and targets for CHES and HWO. Section \ref{sect:DI Model} outlines the methods used to calculate imaging signals in our work. In Section \ref{sect:Improvement}, we discuss the potential enhancements provided by astrometry and introduce a simplified imaging model. The methodology for estimating detection completeness and expected yields of all HWO targets is detailed in Section \ref{sect:Detection}. Section \ref{sect:results} presents the simulation results and an analysis of detection completeness. Finally, we summarize the key findings and provides a discussion of their implications.

\section{Missions and target stars}
\label{sect:Mission}

\subsection{CHES and HWO missions}
This paper investigates the synergistic combination of CHES astrometry and HWO imaging at optical wavelengths. Firstly, CHES is specifically designed to discover terrestrial planets within the habitable zones of approximately 100 nearby solar-type stars \citep{Ji2022, Bao2024, Tan2024}. The CHES satellite, carrying a 1.2m-aperture telescope and 81 scientific CMOS detectors delivering high stability and ultralow distortion, will operate at the Sun-Earth L2 Lagrange point. The astrometric wobble of host stars, induced by Earth-mass planets in 1 au orbits, will be measured with an expected amplitude of about 0.3 $\mu \mathrm{as}$ for a solar-mass star at 10 parsecs, enabling precise detection of such planetary companions. {CHES employs relative astrometry to monitor angular separations between target stars and several distant reference stars over a 5-year baseline, achieving micro-arcsecond-level precision distinct from Gaia's survey, and it adopts a nonuniform sampling strategy that yields period coverage of approximately 60–75\% for a warm planet \citep{Tan2024}}. CHES is expected to detect dozens of terrestrial planets, including Earth analogs, over its duration and will provide orbital elements and mass estimates for these planets \citep{Ji2022}. Additionally, it will accurately measure the planetary masses in the binary systems through RV-astrometry synergy \citep{Huang2025}. Furthermore, CHES allocates about 4100 \textrm{hr} for additional observations, such as nearby HWO targets or compact objects \citep{Tan2024}.

HWO builds upon and integrates concepts from NASA's earlier mission studies, the Habitable Exoplanet Observatory (HabEx) \citep{Gaudi2020} and the Large Ultraviolet Optical Infrared Surveyor (LUVOIR) \citep{LUVOIR2019}, and is scheduled for launch in the 2040s. HWO prioritizes the detection and characterization of potentially habitable planets around nearby stars and the search for biosignatures, using a $\sim 6 $ m space telescope equipped with advanced starlight suppression technologies, preferentially employing a coronagraph, with a potential starshade as a backup plan \citep{Shaklan2023}. As a transformative observatory, HWO enables multidisciplinary breakthroughs in galaxy formation, the evolution of elements, and Solar System studies.

\subsection{Targets Selection}

\begin{figure*}
   \centering
   \includegraphics[scale=0.60]{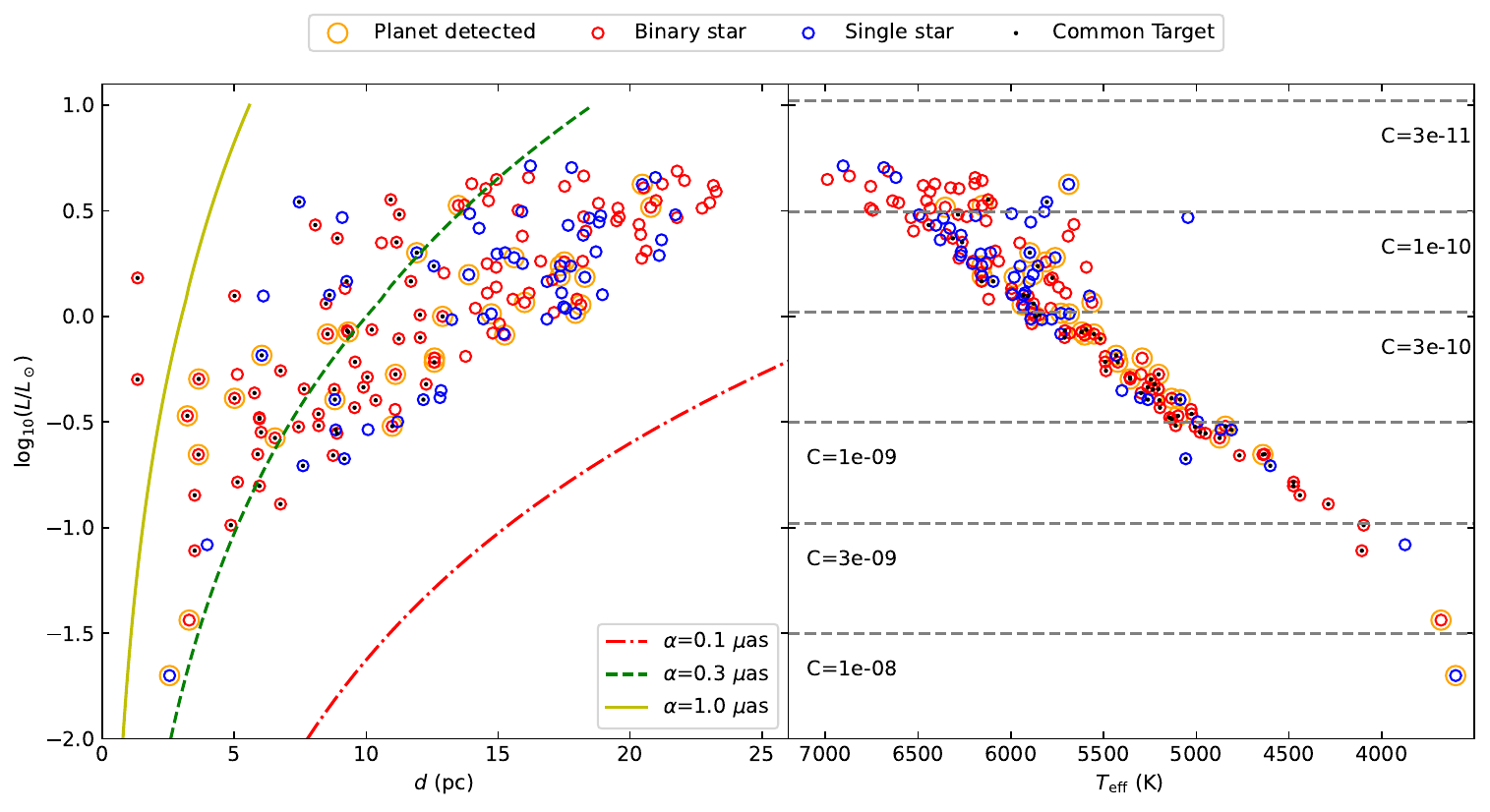}
   \caption{Stellar properties of HWO targets from \citet{Harada2024} and the astrometric and imaging signals contours. The orange, red, blue circle and black points represent stars with known planets, binary stars, single stars and common targets of CHES, respectively. \textit{Left panel}: Stellar luminosity versus distance of all targets. The olive solid, green dashed and red dash-dotted line represent $\alpha=1.0, 0.3, 0.1~\mu \mathrm{as}$ in the scenario of a planet with one Earth mass located at the the center of the habitable zone, the logarithmic relation between stellar mass and luminosity is adopted for drawing the curves \citep{Eker2015}. \textit{Right panel}: Effective temperature versus luminosity, and the horizontal lines corresponds to different imaging contrast.}
   \label{fig:targets}
\end{figure*}

{While HWO's final target list is yet to be determined, we adopt the catalogue filtered from $\sim$ 13,000 nearby stars by \citet{Tuchow2024}, which has also been utilized by \citet{Mamajek2024} and \citet{Harada2024}.} The characteristics of these targets are detailed in \citet{Harada2024}. Here we provide a brief summary: the HWO catalogue includes 164 most accessible main-sequence stars, comprising 66 F-type stars, 55 G-type stars, 40 K-type stars and 3 M-type stars. Based on stellar multiplicity and the existence of circumstellar disks, the targets are classified into three tiers. Tier A includes the top 47 targets with no known complicating factors, while Tiers B and C expand the list by relaxing criteria related to companion separation, brightness, and stellar environments such as binary companions or circumstellar disks. Among them, approximately 60 stars overlap with the CHES targets. Additionally, 30 stars are known to host 70 planets \citep{Kane2024}. It is possible for other stars to host planets based on the planet frequency theory \citep{Hsu2019,Bryson2021}. Moreover, giant planets within habitable zone may be ruled out due to the current radial velocity observations \citep{Kane2024}. The stellar distance, effective temperature, and luminosity are shown in Figure \ref{fig:targets}. The astrometric signal strength ($\alpha$) and imaging contrast ($C$) contours are illustrated in Figure \ref{fig:targets}, as defined in Eqns \ref{equ:astrometric signal} and \ref{equ:contrast}.

\section{Methods of calculating reflected light} \label{sect:DI Model}

\subsection{Orbital geometry}

Considering a system consisting of a star and a planet, located at a distance $d$ from the observer, with the planet's mass and radius denoted as $M_p$ and $R_p$. The planet is assumed to follow Keplerian motion orbiting its host star. The orbit is characterized by six orbital elements: the semi-major axis $a$, eccentricity $e$, inclination $I$, longitude of the ascending node $\Omega$, argument of periapsis $\omega$, and true anomaly $f$, which varies with time. The argument  $\theta$ is defined as $\theta = \omega + f$. The orbital period $P$ is given by Kepler's Third Law.

The distance between the planet and the star, $r$, can be expressed as:
\begin{equation} \label{equ:planet distance}
   r = \frac{a (1 - e^2)} {1+ e \cos f} \ ,
\end{equation}
at a given time $t$, relative to a reference time $t_0$, the true anomaly $f$ can be derived from the eccentric anomaly $E$:
\begin{equation}
\begin{aligned}
   E - e \sin E &= 2 \pi / P \cdot (t - t_0)  \\
   \tan\left(\frac{f}{2}\right) &= \sqrt{\frac{1 + e}{1 - e}} \cdot \tan\left(\frac{E}{2}\right).
\end{aligned}
\end{equation}

Then, the position of the planet in the observed plane can be calculated using the following equations:
\begin{equation}\label{eq5}
\begin{aligned}
   x (t) &= A X(t) + F Y(t) \, , \\
   y (t) &= B X(t) + G Y(t) \, ,
   \end{aligned}
\end{equation}
where the parameters in these equations are derived from the Thiele-Innes equations \citep{Thiele1883}:
\begin{equation}\label{equ:TI}
   \begin{aligned}
   \left\{
       \begin{array}{ll}
       &A=  a (\cos \Omega \cos \omega - \sin \Omega \sin \omega \cos i)  \, , \\
       &B=  a (\sin \Omega \cos \omega + \cos \Omega \sin \omega \cos i)  \, , \\
       &F=  a (-\cos \Omega \sin \omega - \sin \Omega \cos \omega \cos i) \, , \\
       &G=  a (-\sin \Omega \sin \omega + \cos \Omega \cos \omega \cos i) \, , \\
       &X (t) = \cos E(t) - e  \, , \\
       &Y (t) = \sqrt{1-e^2} \sin E(t)  \, . \\
       \end{array}
   \right.
   \end{aligned}
\end{equation}

Thus the projected angular separation $\rho$ can be calculated as:
\begin{equation} \label{equ:rho}
   \rho = \sqrt{x^2 + y^2} / d ,
\end{equation}
Alternatively, $\rho$ can also be determined using geometry relationship \citep{Greco2015}:
\begin{equation} \label{equ:rho2}
   \rho = \left( \sqrt{1 - \sin^2{I}\sin^2\theta }\right) \frac{r}{d}.
\end{equation}

As the planet orbits its host star, the star also undergoes a wobble around their common center of mass due the planet's gravitational influence. The amplitude of this stellar wobbles, $\alpha$, induced by the planet is given by \citep{Perryman2014}
\begin{equation} \label{equ:astrometric signal}
   \alpha = 3.0  \left(\frac{M_p} {M_\oplus}\right) \left(\frac{M_*} {M_\odot}\right)^{-1} \left(\frac{a} {1~\mathrm{au}}\right) \left(\frac{d} {1~\mathrm{pc}}\right)^{-1} \mu \mathrm{as} \ .
\end{equation}

\subsection{Planetary Photometry}\label{sect:contrast}
The reflected light from a planet is a critical quantity observed in direct imaging, as it determines the flux ratio between the planet and its host star, commonly referred to as the contrast. This contrast depends on the planet's albedo, size and phase, and can be expressed by the following equation:
\begin{eqnarray}\label{equ:contrast}
   C = A_g \  \phi (\beta) \left(\frac{R_p}{r}\right) ^2 ,
\end{eqnarray}
where $A_g$ is the geometric albedo of the planet, $\phi(\beta)$ is the phase function parameterized by the phase angle $\beta$, and $r$ is the distance between the planet and the star in Eqn \ref{equ:planet distance}. In practice, the phase function can be highly complex. As a simplification, the study uses the Lambert phase function, which is expressed as \citep{Sobolev1975}:
\begin{equation}
   \phi(\beta) = \left[\sin \beta + (\pi - \beta) \cos \beta \right] / \pi \ .
\end{equation}

The phase angle $\beta$ is defined as the angle between the planet-star vector and the planet-observer vector. Assuming the planet follows Keplerian motion, the phase angle can be derived from $ \cos\beta = \sin{I} \sin \theta$. As the phase angle changes over time, and $r$ also varies in an elliptical orbit, the contrast $C$ can fluctuate. These variations depend on the orbital properties of the planet and its alignment with respect to the observer. Using the planet contrast, the apparent magnitude at V-band of the planet, $m_p$, can be calculated as:
\begin{eqnarray}\label{equ:magP}
   m_p = m_V - 2.5 \ \log_{10} C .
\end{eqnarray}
where $m_V$ is the apparent magnitude of the host star in the V-band.

\section{Improvement from astrometry}
\label{sect:Improvement}

As the planet moves along its orbit around the host star, its magnitude and contrast vary over time. Additionally, portions of the orbits in some cases may fall within the undetectable region near the star due to limitations of the inner working angle (IWA). With the ultra high astrometric precision of CHES, HWO can significantly enhance planet detectability through optimally timed observations.

\subsection{Planetary signal and orbital architectures}

The planetary contrast is observed to vary over time as illustrated in Section \ref{sect:contrast}. The phase angle and planetary distance, derived from Eqn \ref{equ:planet distance}, are directly related to the orbital configuration. This relationship allows us to formulate an optimal strategy to observe the strongest planetary reflected light. By leveraging astrometric data, when the orbital elements are known, the timing of observations can be optimized to maximize visibility.

\begin{figure}
   \centering
   \includegraphics[scale=0.5]{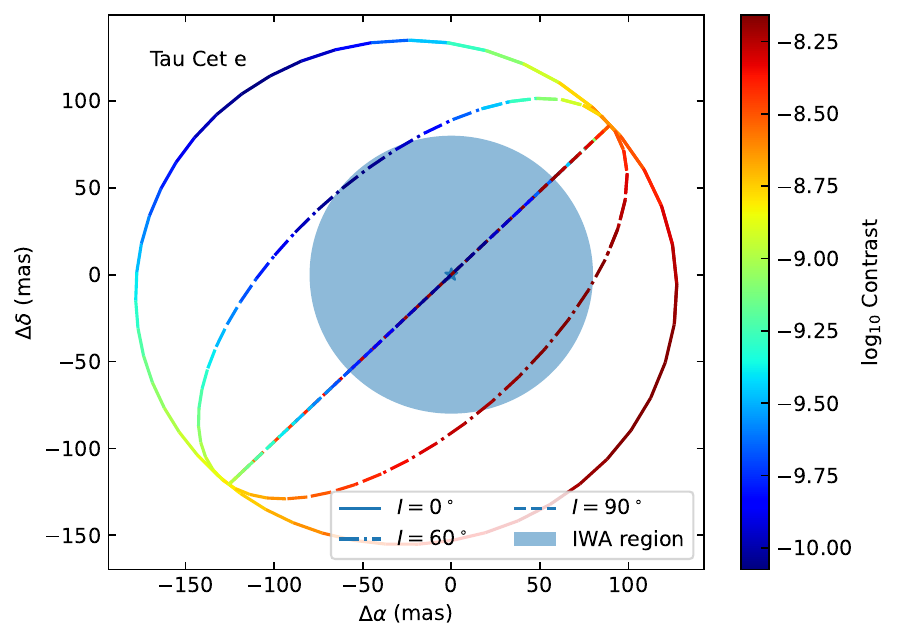}
   \caption{Simulated motion and contrast of $\tau$ Cet e for 200 days. The solid, dashed and dashed-dotted lines represent planet's position for $I =0^\circ$, $60^\circ$, and $90^\circ$, respectively. The colorbar indicates logarithmic contrast, while the blue region marks area within the IWA, where detection is not possible.} \label{fig:taucet}
\end{figure}

We take the planet $\tau$ Cet e as an example, a super Earth orbiting the G8.5 V star $\tau$ Cet, which is a common target of both CHES and HWO. Discovered through radial velocity, $\tau$ Cet e has a mass of approximately 3.9 $M_\oplus$ and an orbital period of 163 days, placing it within the habitable zone \citep{Feng2017}. Since the inclination of $\tau$ Cet e is unknown, we generate simulated observations at three inclinations ($0^\circ$, $60^\circ$, and $90^\circ$), and calculated the corresponding planetary positions and contrast. Figure \ref{fig:taucet} illustrates the simulated motion and contrast of $\tau$ Cet e. The color bar indicates the logarithmic contrast, while the blue region marks the area within the IWA, where detection is not possible. Although the angular semi-major axis of $\tau$ Cet e ($\sim 150\  \mathrm{mas}$) is much larger than the IWA (58 \textrm{mas}), portions of the orbit remain undetectable, particularly for the edge-on orbit.

\begin{figure}
   \centering
   \includegraphics[scale=0.65]{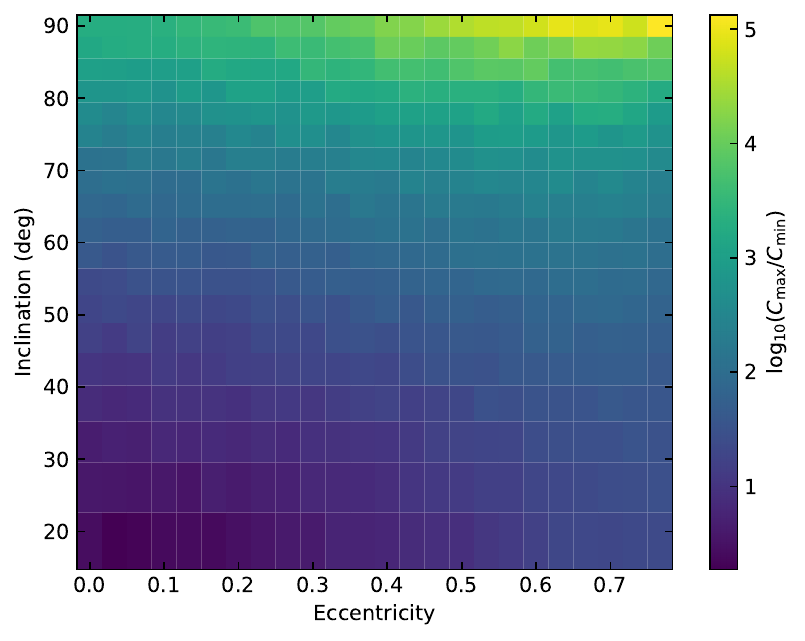}
   \caption{Contrast fluctuation versus with orbital eccentricity and inclination. The eccentricity range is 0 to 0.8, and the inclination range is 0 to 90 degrees. The colorbar presents logarithmic fluctuation between maximum and minimum contrast.}
   \label{fig:con_map}
\end{figure}

As the detectability of a given planet changes over time, it is essential to assess the impact of orbital architectures on detectability. We generate a population of planets with varying orbital elements, and calculate the fluctuation between their maximum and minimum contrast during a single period. The results reveal that contrast fluctuates significantly for planets with high eccentricities and inclinations, consistent with the findings \citep{Kane2013, Greco2015, Kane2018}. As illustrated in Figure \ref{fig:con_map}, the contrast fluctuates by a factor of $10^5$ for planets with high eccentricity and inclination, as discussed by \citet{Greco2015}. This indicates that direct imaging has greater potential for detecting planets when supported by astrometric enhancements. However, as planets become brighter, their angular separation decreases, emphasizing the necessity of high-precision astrometric measurements to observe them at optimal phase windows—capabilities that CHES is designed to achieve. These findings highlight the synergistic potential of combining direct imaging with astrometry for efficient planetary detection.

\subsection{Orbital retrieval and phase prediction}
\label{subsect:retrieval}
The optimal observation phase depends on orbital parameters, which can be derived through the orbital retrieval of CHES \citep{Tan2024}. Here, we provide a brief introduction to the  measurement theory of CHES. The mission employs differential astrometry method to measure the angular separation between the target star and the reference stars. This angular separation varies due to the disparity in proper motion and parallax between the stars, as well as the gravitation influence from planets \citep{Tan2024}. We further assess CHES's capacity to retrieve orbital elements and predict optimal observation phases for HWO targets.

First, we simulate a planet around each target star, assuming an Earth-mass planet located at the center of the habitable zone. We then compute the stellar wobble induced by the planet. Proper motion and parallax are assumed to be removed through our preprocessing and fitting procedure (Tan et al. 2025, in prep). Measurement errors in CHES originate from multiple sources, as discussed in our previous work \citep{Ji2022, Bao2024}. Here, we account for photon noise, telescope noise, detector noise, and stellar activity noise. Detector and telescope noise are modeled as Gaussian noise with standard derivations of 0.36 and 0.74 $\mu \mathrm{as}$, respectively, based on CHES's optical system \citep{Ji2022}. Photon noise is estimated based on stellar magnitude and exposure time. For a target star of magnitude 6 and eight reference stars of magnitude 13, the standard deviation of photon noise after a 2-hour exposure is 0.35 and 0.58 $\mu \mathrm{as}$, respectively \citep{Ji2022}. For targets observed by both CHES and HWO, we adopt the observational schedules provided by \citet{Tan2024}, which specify the uneven observation time series and exposure durations for each star. For targets exclusive to HWO, we assume a uniform observation strategy with 40 observations over a 5-year period and 1-hour exposure time per observation. Additionally, the stellar photocenter shifts due to active regions on its surface as the star rotates. Following our previous work, we simulate spots and faculae based on stellar chromospheric activity indices $\mathrm{log}R^{\prime}_{\mathrm{HK}}$ and estimate the resulting photocenter jitter \citep{Bao2024}. The total noise, approximately 1 $\mu \mathrm{as}$, combines Gaussian noise (photon, telescope, and detector noise) with stellar activity noise. As noted by \citet{TheiaCollaboration2017} and \citet{Perryman2014}, the precision of CHES is sufficient to detect terrestrial planets with just a few dozen observations.

\begin{figure}
   \centering
   \includegraphics[scale=0.5]{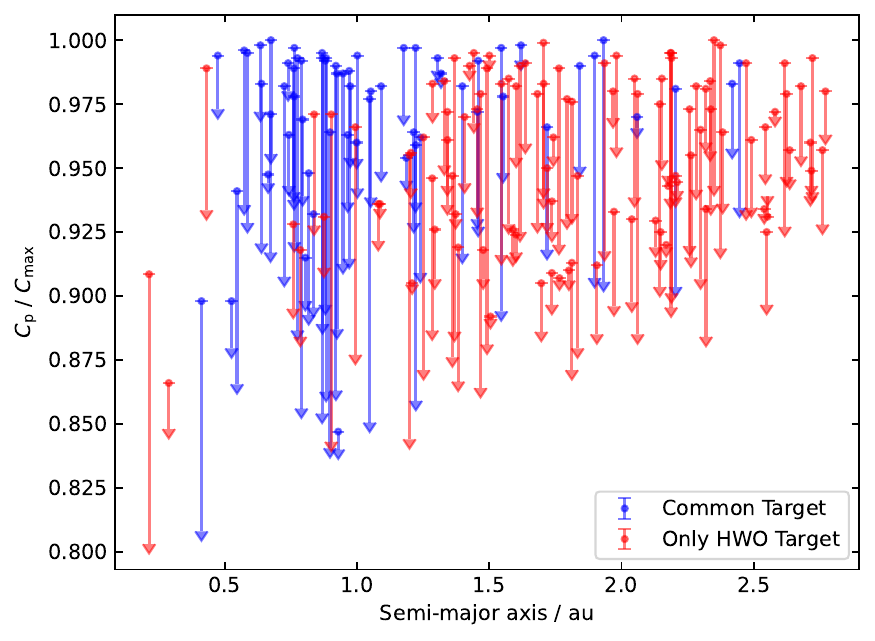}
   \caption{The ratio between $C_p$ and $C_\mathrm{max}$ for all stars, as a function of planetary orbital radius. The errorbars represent the low limit of prediction, derived from predicted contrast with errors. Blue points indicate targets observed by both missions, while red points refer  to those exclusive to HWO.}
   \label{fig:predict}
\end{figure}

The orbital elements and corresponding uncertainties of the injected planets around each target are retrieved using Markov Chain Monte Carlo (MCMC) methods implemented in the \texttt{Nii-C} code \citep{Jin2022, Jin2024}. Based on the fitted parameters, we then estimate the planetary contrast for each detected planet. We adopt the Lambert phase function and assume that the planetary radius and albedo are known and constant. In this case, these two parameters affect only the amplitude of the contrast. For Earth-like planets, this assumption may not be strictly valid due to their intrinsic characteristics; however, modeling such complexities is beyond the scope of this work. Here we define $t_p$ as the time when the predicted contrast reaches its maximum, and $C_p$ as the true contrast value at this time. We then compare the maximum true contrast ($C_\mathrm{max}$) with $C_p$. Due to the inherent imperfections of our fitting procedure, we also compute $t_p$ and $C_p$ using orbital elements that include uncertainties. Figure \ref{fig:predict} illustrates the comparison between $C_p$ and $C_\mathrm{max}$ for all stars, versus the distance of their habitable zone, calculated using Eqn \ref{equ:astrometric signal}. The errorbars reflect the uncertainties in the predicted contrast due to errors in the orbital elements. Following this procedure, for each star, we calculate $C_p$ based on fifty sets of orbital elements with errors, and the errorbars represent the worst predictions among them. The blue and red points denote targets observed by both CHES and HWO and those exclusive to HWO, respectively. Our simulations show that, with the astrometric capabilities of CHES, the targets can be observed at phase where planetary contrast exceeds over 80\% of its maximum. We also find that contrast predictions are slightly more accurate for planets with more distant orbits due to their stronger astrometric signals and slower contrast variations.

\subsection{Signal-to-Noise Criterion for Direct Imaging}
\label{sect:Detection efficiency}
Since the instrumental sensitivity depends on the angular separation leading to different signal for planets with different orbits. Here we investigate the detection limit considering the practical instruments and multiple noise sources at the optical band.

Assuming the detector receives photons from planetary reflected light, the signal is contaminated in practice by background illumination. In this work, we consider three main noise components as following: residual starlight ($C_{\mathrm{sr}}$), solar zodiacal light ($C_\mathrm{z}$) and exozodiacal light ($C_{\mathrm{ez}}$) \citep{Stark2014}. These noise sources can dominate the detected signal, depending on stellar characteristics and observing instruments.

Since the specific design of HWO has yet to be finalized, we use the performance of the Hybrid Lyot Coronagraph (HLC) as a template \citep{Savransky2016, Savransky2017, Nemati2020a, Nemati2020b, Keithly2020} \footnote[2]{\url{https://github.com/hsergi/Roman_Coronagraph_ETC}}. The key parameters include three angular-dependent functions: the core contrast $f_{\mathrm{con}}$, which characterizes the coronagraph's suppression capacity; the core throughput, $f_{\mathrm{thu}}$, which reflects the system's throughput within the photometric aperture; and the coronagraphic transmission, $f_{\mathrm{tra}}$, which quantifies the intensity transmission of extended background sources. {We utilize the original instrumental data provided by \texttt{RCETC} \citep{Hildebrandt2023ascl}, which are subsequently normalized to the baseline specifications ($\mathrm{con}_0$, $\mathrm{thu}_0$, and $\mathrm{tra}_0$) for HWO as listed in Table~\ref{tab:optical-input} \citep{NAS2021}.}

\begin{figure}
   \centering
   \includegraphics[scale=0.45]{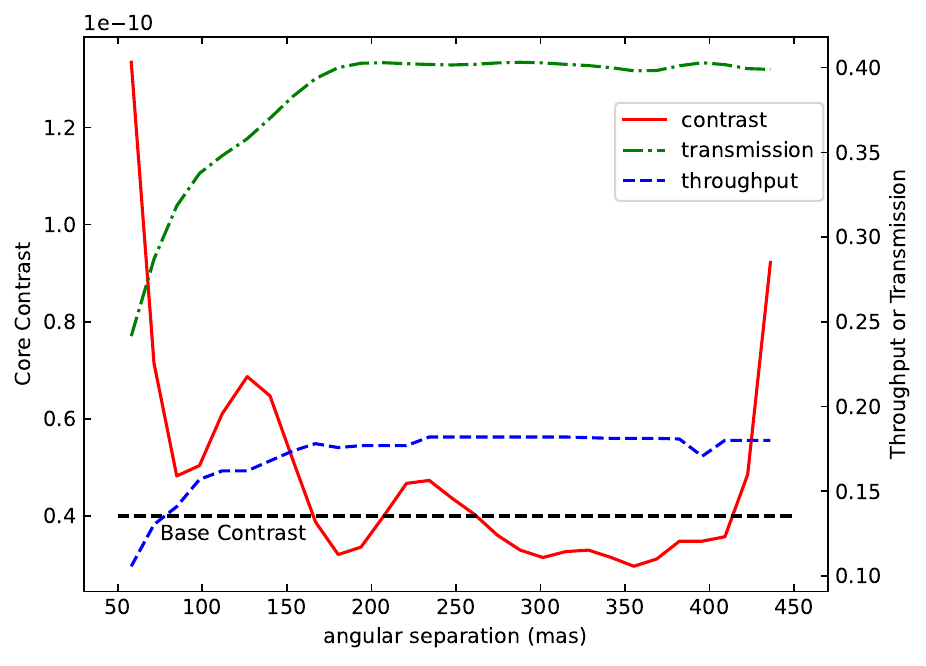}
   \caption{Instrumental performance curves used in our simulations. The red solid, blue dashed and green dashed-dotted lines represent core contrast, core throughput and coronagraphic transmission, respectively. The black lines shows a base contrast value $4\times 10^{-11}$ of HWO.}
   \label{fig:inst}
\end{figure}

Firstly, the original minimum anglual separation of 155 \textrm{mas} was linearly reduced to the IWA in Table \ref{tab:optical-input}. While the outer working angle (OWA) of HWO has not yet been finalized, the National Academies of Sciences, Engineering, and Medicine estimates an approximate OWA of 1000 \textrm{mas} \citep{NAS2021}, whereas \citet{Mennesson2024} suggest a value around 300 \textrm{mas}. Since nearly all the considered planets (except those near the outer boundary of the habitable zone of Alpha Cen A and B) lie within HLC's maximum angular separation of 463 \textrm{mas}, we retain this value as the outer boundary and assume that performance remains constant beyond this value. Next, the contrast, efficiency and transmission of HLC are linearly adjusted so that their median values align with the baseline specifications ($\mathrm{con}_0$, $\mathrm{thu}_0$, and $\mathrm{tra}_0$) in Table \ref{tab:optical-input}. {Following the HabEx design \citep{Gaudi2020}, our simulations incorporate a single coronagraph with a 20\% bandwidth across the passband. Figure~\ref{fig:inst} illustrates the variation of these parameters with angular separation. A summary of all parameters used in the analysis is provided in Table~\ref{tab:optical-input}.}

The stellar leakage is a major source of noise in direct imaging, primarily due to the imperfections in the instruments. We estimate the residual light for a prototype coronagraph as follows. First, the star radiation is assumed to follow a blackbody spectrum. The flux $F_{\mathrm{st}}$ is calculated over the given bandpass in our simulations, and the stellar photon count rate is estimated using the equation:
\begin{equation}\label{equ:cst}
   C_{\mathrm{st}} =   q \cdot a_0 \cdot F_{\mathrm{st}} \cdot (\pi D^2 /4)  ,
\end{equation}
where the description and value of parameters are defined in Table \ref{tab:optical-input}.

Next, the residual starlight and planetary photon count rate after passing the coronagraph are calculated as \citep{Delacroix2016}:
\begin{equation}\label{equ:count_sr_pl}
   \begin{aligned}
      C_{\mathrm{sr}} &= C_{\mathrm{st}} \ f_{\mathrm{con}}(\rho) \ f_{\mathrm{thu}} (\rho)  \\
      C_{\mathrm{pl}} &= C \ C_{\mathrm{st}} \ f_{\mathrm{thu}} (\rho)
      \end{aligned}
   \end{equation}
where the $C$ and $\rho$ is planet's contrast and angular separation, respectively.

The local zodiacal light is another major noise source in direct imaging. Its intensity depends on the observational orientation relative to the Sun. For our simulations, we assume that HWO is located at the Sun-Earth L2 point \citep{Mamajek2024}. The observatory's coordinates are acquired from JPL Horizons\footnote[1]{\url{https://ssd.jpl.nasa.gov/horizons/app.html}}. The \texttt{PyMsOfa} package is used to compute the vector positions of observatory and the Sun \citep{Ji2023}.
The intensity of zodiacal light, $I_{\mathrm{zod}}^{\lambda}$, at a given orientation is interpolated from observational data by \citet{Leinert1998}, taken at 500 nm. {Assuming a target with a relative ecliptic longitude of $75^\circ$ and latitude of $45^\circ$, $I_{\mathrm{zod}}^{\lambda}$ is approximately $3\times10^4  \  \mathrm{ph \ / (cm^2~\mathrm{s}~sr~\text{\AA})}$  \citep{Leinert1998}.} The photon count rate from zodiacal light is calculated as \citep{Stark2014, Delacroix2016}:
\begin{equation}
 C_\mathrm{z} =  q \cdot a_0 \cdot \pi \left(\frac{ D}{2}\right)^{2} \cdot f_{\mathrm{tra}}(\rho) \cdot \frac{\pi}{2} \left(\frac{ \lambda}{D}\right)^{2}   \cdot I_{\mathrm{zod}}^{\lambda} \cdot \Delta \lambda.
\end{equation}
The brightness of exozodiacal light can be calculated as:
\begin{equation}
  F_{\mathrm{ez}} = N_{\mathrm{ez}} \cdot 10^{-0.4 (M_{\mathrm{ez}}+m_{V} - m_{\odot})} \cdot f(I) \cdot \left( \frac{r}{1~ \mathrm{au}}\right)^{-2} \mathrm{arcsec}^{-2},
\end{equation}
where $N_{\mathrm{ez}}$ is the exozodiacal level defined by \citet{Stark2014}, we adopt $N_{\mathrm{ez}}=3$, a median value derived from the HOSTS survey observations \citep{Ertel2020}, $m_{V}$ and $m_{\odot}$ are the magnitudes at V band of target star and the Sun, respectively. $M_{\mathrm{ez}}$ is the surface brightness of the exozodiacal dust at 1 au, adopted as $M_{\mathrm{ez}}=22 \ \mathrm{mag\  arcsec^{-2}}$ \citep{Turnbull2012}. $f(I)$ is a dimensionless factor that depends on planetary inclination \citep{Savransky2010, Stark2014}, and $r$ is the planet-star distance given by Eqn \ref{equ:planet distance}.

{The photon count rate from exozodiacal light can be estimated using the following equation:}
\begin{equation}\label{equ:cez}
   C_{\mathrm{ez}} = q \cdot a_0 \cdot {\pi}\left(\frac{ D}{2}\right)^{2} \cdot F_{\mathrm{ez}} \cdot f_{\mathrm{thu}}(\rho) \cdot \frac{\pi}{2} \left(\frac{ \lambda}{D}\right)^{2}  \cdot F_0  ,
 \end{equation}
{where $F_0$ is the zero-magnitude flux integrated over the given bandpass, estimated from the Vega spectrum, with a value of approximately $1.2\times 10^6~\mathrm{ph}/(\mathrm{cm}^2~\mathrm{s})$ \citep{Stark2014, Traub2016, Delacroix2016}. The calculations are performed using the \texttt{synphot} package \citep{synphot2018}}.
The signal-to-noise ratio for a planet in direct imaging is defined as follows, based on \citet{Stark2014}:
\begin{equation}
   \mathrm{S/N} = \sqrt{\frac{ C_{\mathrm{pl}}^2}{C_{\mathrm{pl}}+2 C_\mathrm{b}} \tau} .
\end{equation}
where $C_{\mathrm{pl}}$ is photon count rate from the planet, $C_b$ is photon count rate from background, which includes the residual starlight, zodiacal and exozodiacal light, and dark current (as defined in Table \ref{tab:optical-input}), and $\tau$ is the exposure time. {The number of pixels is set to eight in accordance with the Nyquist sampling criterion to ensure adequate spatial resolution \citep{Gaudi2020, LUVOIR2019}.}

\section{Detection completeness and Expected Yields}
\label{sect:Detection}
\subsection{Detection completeness without prior knowledge}
\label{subsect:DetectionDI}
To evaluate the impact of orbital architectures on detectability and signal strength, we estimate the detection completeness, for each star in HWO catalogue. The detection completeness $C_\mathrm{det}$ is defined as which is the fraction of possible planets that will be detected for a given star \citep{Brown2004}, and is calculated using the criteria (the parameters listed in Table \ref{tab:optical-input}) outlined in \citet{Kasdin2006}, \citet{Mamajek2024} and \citet{Mennesson2024}:
\begin{itemize}
   \item The planet's angular separation must lie between the OWA and IWA: $\mathrm{OWA}>\rho>\mathrm{IWA}$
   \item The planet's magnitude must exceed the magnitude limit: $m_p < m_{\mathrm{floor}}$
   \item The signal-to-noise (S/N) ratio exceed the threshold: $\mathrm{S/N}>7$
\end{itemize}
{Remarkably, we adopt a more practical $\mathrm{S/N}$ criterion \citep{Kasdin2006, Mennesson2024} rather than simply considering a planet detectable if its contrast exceeds the detection limit (i.e. $\mathrm{con}_{\mathrm{0}}$ of HWO). Additionally, although JWST has demonstrated observations interior to the IWA, we restrict our analysis to detections exterior to the IWA, following \citet{Greco2015} and  \citet{Plavchan2024}. This simplification is justified by our primary focus on astrometric enhancements.}

\begin{table*}
   \begin{center}
   \caption{Optical model parameters} \label{tab:optical-input}

    \begin{tabular}{clclc}
   \hline\noalign{\smallskip}
   \hline\noalign{\smallskip}
   Parameter  & Symbol &  Value/range & Unit& Reference \\
   \hline
      Aperture    & $D$ &     6  & \textrm{m}   &    (a)  \\
      Wavelength  &  $\lambda$      & 500   &  nm  &   (b)  \\
      Band width               &  $\Delta \lambda$   & 100       &  nm   &  (b)   \\
      Quantum efficiency   &  $q$     & 0.9    &     &  (c) \\  
      Optics attenuation    & $a_0$ &     0.5  &    &     \citet{Delacroix2016}\\
      Inner working angle     & IWA   &   58    & \textrm{mas}  & (b, d) \\
      Outer working angle     & OWA   &   1000    & \textrm{mas}  & (d)\\
      Magnitude limit      & $m_{\mathrm{floor}}$     & 31   & mag & (a) \\
      Base core contrast     & $\mathrm{con}_0$   &   $4\times 10^{-11}$    &  & (a) \\
      Base core throughput     & $\mathrm{thu}_0$   &    0.18   &   & (a) \\
      Base coronagraphic transmission     & $\mathrm{tra}_0$  &  0.4    &   &  \citet{Mennesson2024} \\
      {Dark current}  &  $\xi$     & $3\times 10^{-5}$    &  {$\rm{e}^{-1} \ \rm{pix}^{-1} \ s^{-1}$}   &  (c)   \\
   \hline
   \hline
   \end{tabular}
   \end{center}
   \tablecomments{Reference: (a) \citet{Mamajek2024}, (b) \citet{Gaudi2020}, (c) \citet{Stark2024}, (d) \citet{NAS2021}. }
\end{table*}

To estimate the completeness, we inject a population of planets around each target, with planetary parameters randomly generated according to the specifications in Table \ref{tab:planet}. The planetary albedo is randomly sampled between 0.2 and 0.4. The upper bound is based on the estimation of rocky planets at optical band by \citet{Mallama2017}, while the lower bound is provided by \citet{Stark2014}, as the Rayleigh scattering effects are diminished for planets around cooler stars \citep{Mamajek2024}. The detection completeness $C_\mathrm{det}$ is then calculated based on the above criteria in the following paragraph. In this analysis, a habitable planet is considered to have a radius between $0.5 \sim 1.5~R_\oplus$ \citep{Kammerer2022}, and its semi-major axis is derived from the methods described by \citet{Kopparapu2014}. The inner and outer edges of the habitable zone are calculated based on the stellar effective temperature and luminosity. {Compared with the circular orbits assumption \citep{Stark2015,Morgan2021, Plavchan2024}, we incorporate eccentric orbits in our analysis, as the planetary contrast in strongly influenced by eccentricity as shown in Figure \ref{fig:con_map}. Additionally, planetary detectability is likewise affected by eccentricity \citep{Greco2015}.}

\begin{table*}
   \begin{center}
   \caption{Planetary parameters} \label{tab:planet}
    \begin{tabular}{lclch}
   \hline\noalign{\smallskip}
   \hline\noalign{\smallskip}
   Parameter  & Symbol &  Input distribution & Unit& Distribution \\
   \hline
      Albedo         & $A_g$  & $ \mathcal{U}(0.2, 0.4)$  & & \\
      Semi-major axis  &  $a$      & habitable $\mathrm{zone}^{*}$   &  au  &  Uniform   \\
      Eccentricity     &  $e$   &  $ \text{Beta}(1.12, 3.09)^{**} \ $      &     &  Beta    \\
      Radius  &  $R_{p}$      & $ \mathcal{U}(0.5, 1.5)$   &  $R_{\oplus}$  &  Uniform   \\
      Cosine inclination  &   $\cos I$ &$ \mathcal{U}(0, 1)$  &    & Uniform for Cosine \\
      Argument of perihelion  &  $\omega$     & $ \mathcal{U}(0, 360)$   &  deg  &  Uniform   \\
      Initial true anomaly  &  $f_0$      & $ \mathcal{U}(0, 360)$   &  deg  &  Uniform   \\
   \hline
   \end{tabular}
   \end{center}
   \tablecomments{(*) The habitable zone range is based on model of \citet{Kopparapu2014}. (**) The parameters of distribution are taken from \citet{Kipping2013}.}
\end{table*}

To evaluate the detection completeness using direct imaging alone, {we assume that for a specific star, the imaging without prior knowledge occurs $N_\mathrm{v}$ times, with an interval $\Delta T$ between observations and a single exposure time $\tau$. As described by \citet{Stark2015}, these parameters can be optimized to maximize completeness while saving observation time.}

To determine appropriate values for these parameters, we generated a set of planets within the conservative habitable zone and attempted to recover them based on the established criteria. The parameter space is set as follows: $N_\mathrm{v}$ ranges from one to six \citep{Mamajek2024}, $\Delta T$ spans from ten days to half of the period of a planet located at the center of the habitable zone \citep{Stark2015}, with 20 uniform samples in this range; and $\tau$ varies from one to ten hours in 1-hour increments (we find that completeness increases only marginally with longer exposure times).

The completeness are then ranked alongside their corresponding time requirements ($N_\mathrm{v} \times \tau$). The parameters combination yielding the highest completeness is selected unless the difference between this value and the next highest value is less than 2\%, in which case the latter is chosen to optimize time efficiency.

Additionally, we assess detection efficiency using a benefit-to-cost ratio \citep{Stark2015}, defined as
\begin{equation}\label{equ:benefitDI}
   f_\mathrm{DI} = \frac{\overline{\mathrm{S/N}} \ C_\mathrm{det}}{N_\mathrm{v} \times \tau},
\end{equation}
where $\overline{\mathrm{S/N}}$ represents the mean $\mathrm{S/N}$ of detected planets.

{Compared with the definitions by \citet{Stark2015}, our formulation explicitly incorporates $\mathrm{S/N}$, as a higher $\mathrm{S/N}$ reduces the false-positive rate, and enhance the quality of simultaneous spectrum observations. Stellar parameters are sourced from the HWO catalog by \citet{Harada2024}, allowing us to derive the optimal observational parameters for each target. }

\subsection{Detection completeness with prior knowledge}

To explore the improvements from astrometric assistance, we assume that target stars are observed as the optimal phase predicted from prior astrometric data. Several studies investigated the benefits of prior knowledge for planet yields and detection efficiency in direct imaging. \citet{Davidson2011} and \citet{Morgan2021} show that the observation schedules can be optimized using exoplanet priors from astrometry and extreme precision radial velocity (EPRV), respectively. In addition, \citet{Plavchan2024} derive analytic expressions in telescope diameter and detection efficiency based on perfect precursor knowledge in single-planet system.

Since nearly half of known planets reside in multi-planet systems and eccentric orbits, and some seemingly solitary planets may have undetected companions, such system are particularly valuable for direct imaging observation. Observing multiple planets within the same system can be more time-efficient and scientifically rewarding. Here we extend CHES's capabilities into two-planet systems and incorporate them into our analysis. We modified the completeness calculating methods from \citet{Stark2015} to account for synergy observations. In our simulations, we randomly generate systems containing one or two planet using parameters listed in Table \ref{tab:planet}, assuming their orbits are already known. Then number and timing of observations are then optimized. For two-planet systems, we determine whether to observe both planets simultaneously or separately by maximizing the benefit-to-cost ratio.

Firstly, we consider observe two planets (denoted as a and b) simultaneously. Using the methods described in Sect \ref{subsect:retrieval}, we calculate the planetary contrast at the phase where the contrast reaches its maximum while ensuring the planet is exterior to the IWA, and the uncertainties of predicted contrast are considered as shown in Figure \ref{fig:predict}. The exposure time, $\tau_s$, is chosen such that the $\mathrm{S/N}>7$ are satisfied for both planets, ensuring that both planets meet the magnitude and IWA criteria. We then compute the benefit-to-cost ratio for this system as $f_1 = \left({\mathrm{S/N}_a+\mathrm{S/N}_b}\right)/{\tau_s}$.

However, simultaneous observations may not always feasible due to orbital architectures. In such cases, we consider individual observations. We optimize the observations phase and exposure time, $\tau_a$, $\tau_b$, to maximize the benefit-to-cost ratio $f_2 = \left({\mathrm{S/N}_a+\mathrm{S/N}_b}\right)/\left({\tau_a + \tau_b}\right)$. Finally, we compare $f_1$ and $f_2$ to determine the preferred observation strategy for this system. In some cases, neither approach enables the detection of both planets due to instrumental limit. And the completeness with astrometry is also estimated. Following the approach in Sect \ref{subsect:DetectionDI}, we also calculate the overall benefit-to-cost ratio for synergy observations, as:
 \begin{equation}\label{equ:benefitDA}
   f_\mathrm{DI+AST}= \dfrac{\overline{\mathrm{S/N}} \ C_\mathrm{det}}{\overline{{N_\mathrm{v} \times \tau}}} ,
\end{equation}
where is $\overline{{N_\mathrm{v} \times \tau}}$ is the average time consumption across all simulated systems. It is worth noting that $f_1$ and $f_2$ apply to individual systems, whereas $f_\mathrm{DI}$ and $f_\mathrm{DI+AST}$ are computed over large planetary populations.

\begin{figure*}
   \centering
   \gridline{\fig{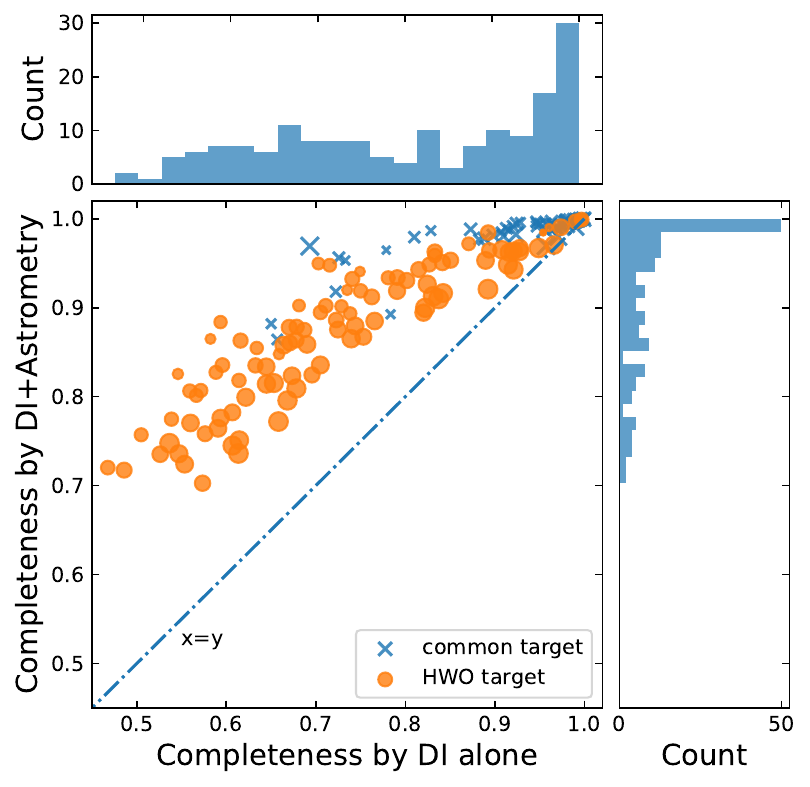}{0.43\textwidth}{(a)}
   \fig{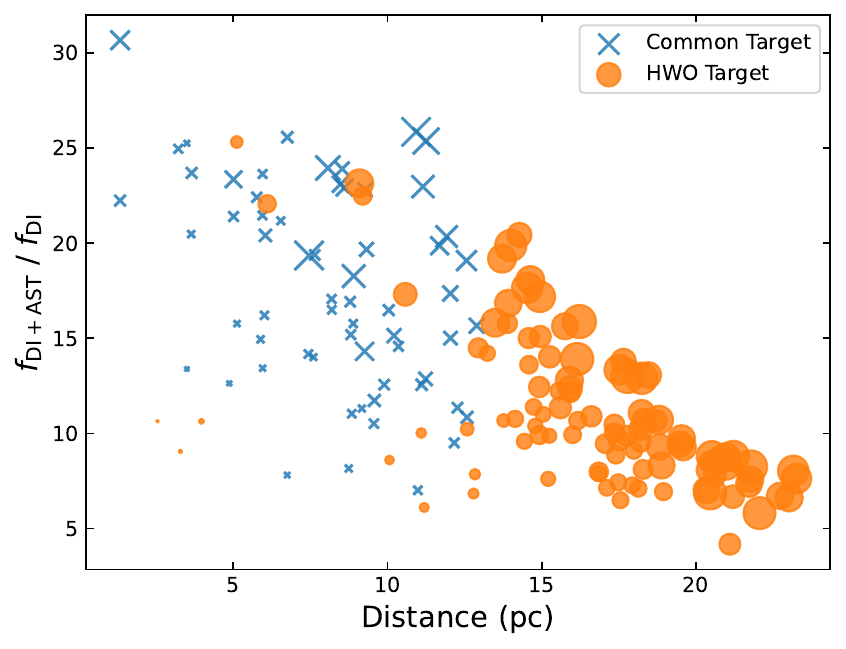}{0.45\textwidth}{(b)}}
   \caption{\textit{Left panel}: The overall completeness of every target. The bottom-left sub-panel shows comparison of completeness by direct imaging alone (x-axis) and imaging+astrometry (y-axis). The top-left and bottom-right sub-panels give histogram of completeness in two scenarios. \textit{Right panel}: The comparison of benefit-to-cost ratio by imaging alone and imaging with astrometry. The orange points and blue crosses indicate targets for HWO only and common targets, respectively. The size of points represent the distance of the center of habitable zone. }
   \label{fig:compDI}
\end{figure*}

Figure \ref{fig:compDI} presents the estimated completeness and improvements in benefit-to-cost ratio for planets within the conservative habitable zone of each target, comparing direct imaging alone and imaging with astrometric assistance. The size of the points represents the distance of habitable zone, highlighting that stars with closer habitable zones tend to yield higher completeness due to stronger planetary contrast. Additionally, we also find the completeness decreases with stellar distance. For most targets, completeness does not reach 100\% due to instrumental limits. However, we find that incorporating astrometry increases completeness by $\sim$ 10\%, with our  results indicating that the improvements are more pronounced for more distant stars. Additionally, the benefit-to-cost ratio is enhanced by a factor of two to {thirty} times compared to cases without prior knowledge, depending on the stellar distance. Notably, the improvements in detection efficiency are most pronounced for nearer stars, which are also key targets for both of CHES and HWO.  The detailed results for each target are listed in Table \ref{tab:yield_all}.

\begin{figure*}
   \gridline{\fig{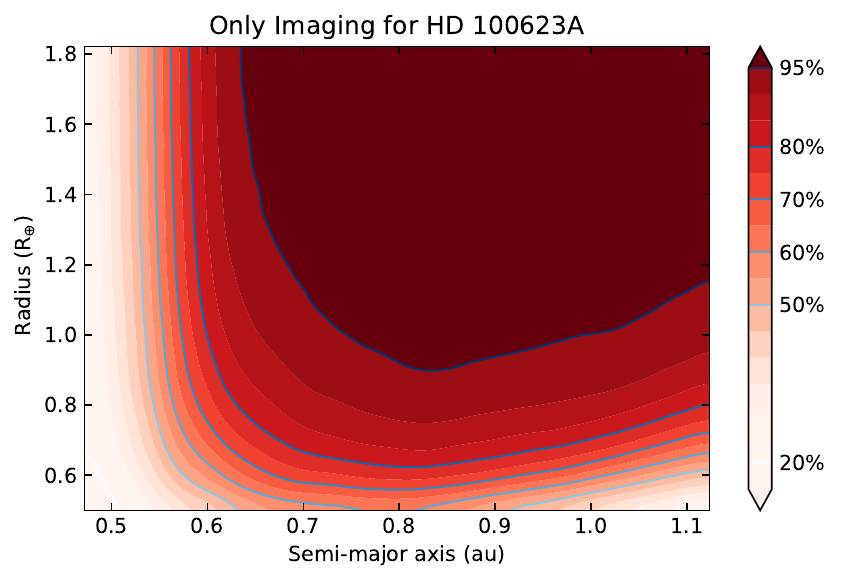}{0.45\textwidth}{(a)}
            \fig{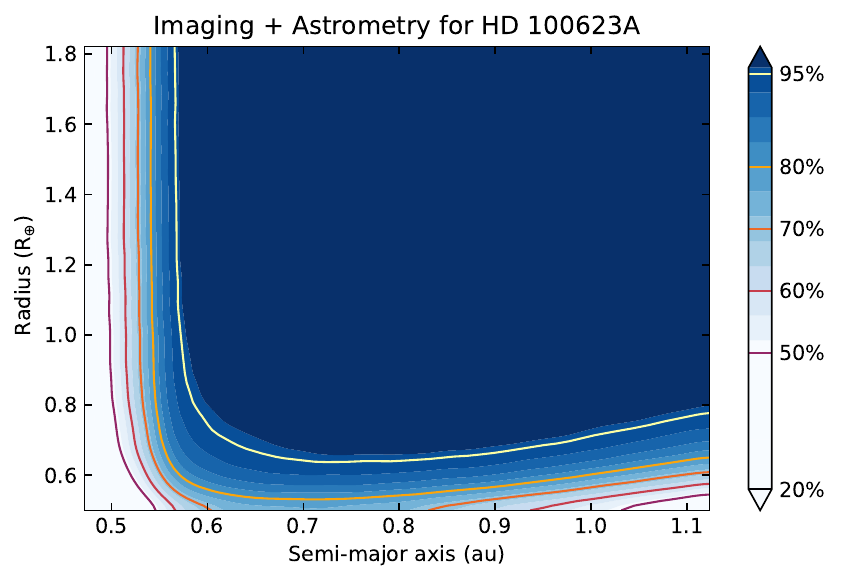}{0.45\textwidth}{(b)}}
   \caption{The detection map for HD 100623A in two observational scenarios. \textit{Left panel}: Results based solely on the direct imaging method are derived from five observations, each with a single exposure time of 4~hr. \textit{Right panel}: Results for the synergy between direct imaging and astrometry. Contour lines in different colors represent detection efficiencies ranging from 20\% to 95\%, calculated based on the S/N criterion.}
   \label{fig:detect_map}
\end{figure*}

Since direct imaging sensitivity depends primarily on planetary radius and semi-major axis, we construct girds over $R_p$ and $a$. The planetary radius is set between 0.5 and 1.5 Earth radii, while the semi-major axis is constrained to the optimistic habitable zone by \citet{Kopparapu2014}. We calculate the detection completeness in each cell through methods above. This allows us to create detection maps for both observational scenarios. In the direct imaging-only scenario, the completeness is calculated using the optimal $N_\mathrm{v}$, $\Delta T$ and $\tau$. The results for HD 100623 A are shown in Figure \ref{fig:detect_map}. The left panel demonstrates that the detection completeness for habitable planets using direct imaging alone ranges from 20\% to 95\%. With the assistance of CHES, this completeness increases by approximately 8\%, and the contours shift, allowing the detection of closer-in and smaller planets.

\subsection{Planetary Occurrence Model}

The goal here is to estimate the number of planets detectable by HWO, leveraging a planet frequency model. By combining stellar and planetary population statistics with observational constraints, we aim to assess the potential yields of the mission.

The differential population rate model introduced by \citet{Bryson2021} is adopted. Their stellar populations consists of more than 80,000 FGK stars, consistent with targets in our samples. The planet data are drawn from the Kepler DR25 planet candidate catalog, which includes about 4,000 planet candidates with radii in the range $0.5~R_\oplus < R_p < 2.5~R_\oplus$ and instellation flux (i.e., host star flux incident on the planet) in the range $0.2~I_\oplus< I_p <2.2~I_\oplus$. The incident flux is calculated using the effective temperature and planetary semi-major axis, based on the methods by \citet{Kopparapu2014}.

From those samples, \citet{Bryson2021} derived a relationship between occurrence rates $\eta_\oplus$ and stellar or planetary parameters. {They identify a potential dependence of $\eta_\oplus$ on the stellar $T_{\mathrm{eff}}$ attributed to wider habitable zone around hotter stars.} We use their first model which is expressed as:
\begin{equation}
   \frac{\partial^2 N(R_p, I_p, T_{\mathrm{eff}})}{\partial R_p \partial I_p} = F_0 C_0 R_p^\alpha I_p^\beta  T^\gamma  g( T_{\mathrm{eff}}) ,
\end{equation}
where $N$ is the average number of planets per star, $F_0$ is the number of planets per star within the specified range provided by \citet{Bryson2021}, $T=T_{\mathrm{eff}}/{T_\odot}$, is the ratio of stellar and solar effective temperature, $C_0$ is a normalization constant, making the integral of the equation over the specified radius and flux ranges equals $F_0$. Here $g( T_{\mathrm{eff}})$ is a function of the effective temperature $ T_{\mathrm{eff}}$, described in Eqn (4) of \citet{Bryson2021}, which accounts for the broader habitable zones of hotter host stars. $\alpha$, $\beta$, and $\gamma$ are fitted coefficients.

We use the parameters from Poisson likelihood method based on the FGK star samples ($3900~\mathrm{K} \leq T_{\mathrm{eff}} \leq 7300~\mathrm{K}$), which align most stars in our sample. Here we consider three types of habitable planets: the first two types include planets in the optimistic or conservative habitable zones \citep{Kopparapu2014}, with radii $0.5~R_\oplus < R_p <1.5~R_\oplus$ \citep{Kammerer2022}. The third type consists of Earth-like planets, located in the conservative habitable zone and with radii $0.8~R_\oplus < R_p <1.4~R_\oplus$ \citep{Stark2019}.

For three M dwarf stars in our sample (HD 202560, HD 217987, and HD 95753), the previous studies have reported an $\eta_\oplus = 0.33$ \citep{Hsu2020}, consistent with those from Bryson's methods.  Other investigations suggest no significant increase in $\eta_\oplus$ from FGK to M stars \citep{Bryson2021, Bergsten2023}, which contradicts earlier claims that M dwarfs host more planets than FGK stars \citep{Mulders2015, Shields2016}. Accordingly, we assume that the results are applicable to all three stellar types.

\begin{figure}
   \centering
   \includegraphics[scale=0.49]{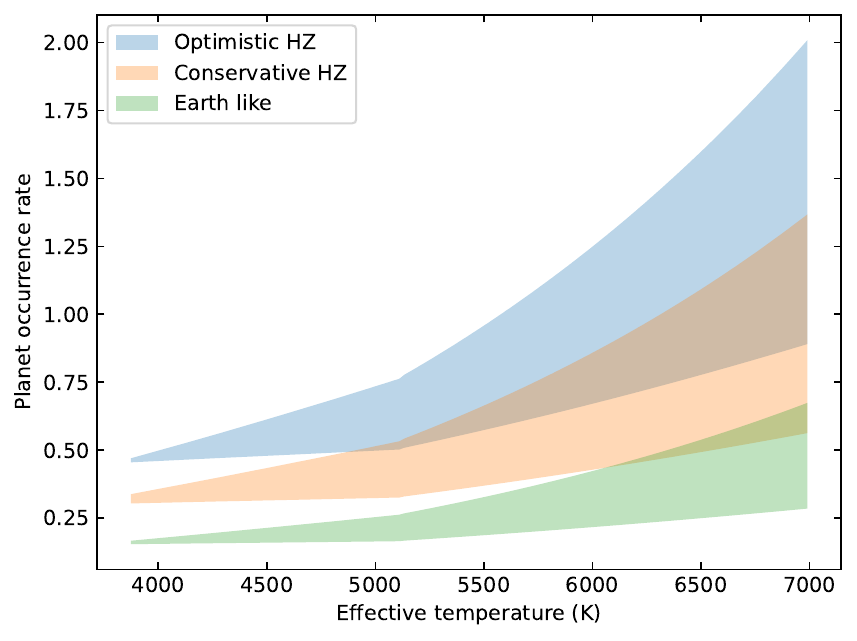}
   \caption{Planet occurrence rate versus corresponding host star effective temperature. The blue, orange and green regions represent occurrence rate of planets in optimistic habitable zone, conservative habitable zone, and Earth-like planets, respectively. Data based on planet occurrence model by \citet{Bryson2021}.}
   \label{fig:teff-occ}
\end{figure}

The occurrence rates for all targets are shown in Figure \ref{fig:teff-occ}. The blue, orange, and green regions represent the rates for planets in the optimistic habitable zone, the conservative habitable zone, and Earth-like planets, respectively. The bounds of occurrence rate for each target are calculated using the low and high model \citep{Bryson2021}. The planet frequency increases with $T_{\mathrm{eff}}$, consistent with those of \citet{Kammerer2022}.

As the methods applied in Section \ref{sect:Detection efficiency}, we make many grids with varying radii and semi-major axes for each target, Then we
compute the occurrence rate $\eta_\oplus$ and detection completeness $C_\mathrm{det}$ within each cell. The planet yields in each cell are calculated $\eta_\oplus \times C_\mathrm{det} $. Finally, the overall yields of detected planets $N_p$ are calculated as the sum of all cells.

\section{Results}
\label{sect:results}
Our findings indicate that the detection completeness and efficiency of direct imaging can be improved through the synergy with future astrometric observations from CHES. Simulations highlight the critical influence of orbital parameters, such as eccentricity and semi-major axis, on detection probabilities. Planets with higher eccentricity exhibit greater variability in contrast and angular separation, which can either enhance or hinder detectability depending on observational timing. Similarly, instrumental parameters of HWO, including the IWA, OWA, and contrast limits, impose stringent constraints on detectable planetary systems.

Using the aforementioned method, {we calculate the completeness following \citet{Stark2015} and extend it to two-planet systems. Our results show that completeness increase slightly ($\sim 10\%$) with the prior astrometric observations. Additionally, the detection efficiency improves greatly by about two to {thirty} times. Then the number of planets expected to be detected are estimated for each HWO target under two observational scenarios: direct imaging alone and imaging combined with astrometry.}

With occurrence rates derived by \citet{Bryson2021} and modeling planetary populations of HWO's target stars, we conclude that HWO could detect about 15 $\sim$ 60 habitable planets, depending on the assumed occurrence rate and the habitable zone model. Table \ref{tab:yield} provides the total number of expected planet yields. {We find that the prior knowledge can enhance yields by $10\% \sim 20\%$ while reducing the number of required observations. \citet{Morgan2021} estimate that planets yields can increase by $\sim 30\%$, with prior radial velocity observations at 3 $\mathrm{cm/s}$ precision. Our enhancements in yields are slightly lower, likely due to more noise sources are considered, and a more conservative approach in optimizing observations.} {Our conservative estimate (based on low occurrence and conservative habitable zone model) is broadly consistent with the science goal of $\sim$ 25 by the Astro2020 Final Report \citep{Mamajek2024}. With prior astrometric observations from CHES, the number could increase slightly by 5 $\sim$ 10, partly due to the eccentric planets in our simulations. Significantly, the completeness with astrometry do not increase considerably ($\sim$ 10\%), because the mission can also be optimized by revisit and appropriate scheduling without any prior knowledge \citep{Stark2014,Stark2015}. While the detection efficiencies will be enhance greatly by up to {thirty} times. Additionally, \citet{Stark2024} use the coronagraph from the LUVOIR-B study as a prototype to estimate HWO yields. Their results also show that the uncertainties of planet occurrence rates dominate the final yield, and the $\mathrm{P}_{25}$ (fraction of the yields larger than 25 terrestrial planets in habitable zone) exceeds 95\%, based on a similar bandpass and instrumentation with slightly lower precision.}

\begin{figure*}
   \centering
   \gridline{
      \fig{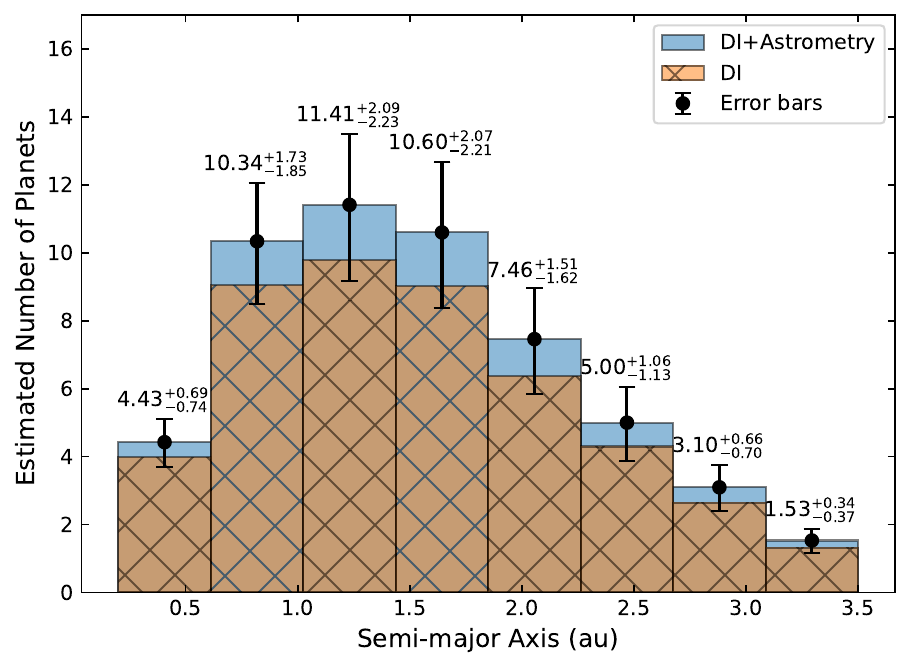}{0.45\textwidth}{(a)}
   \fig{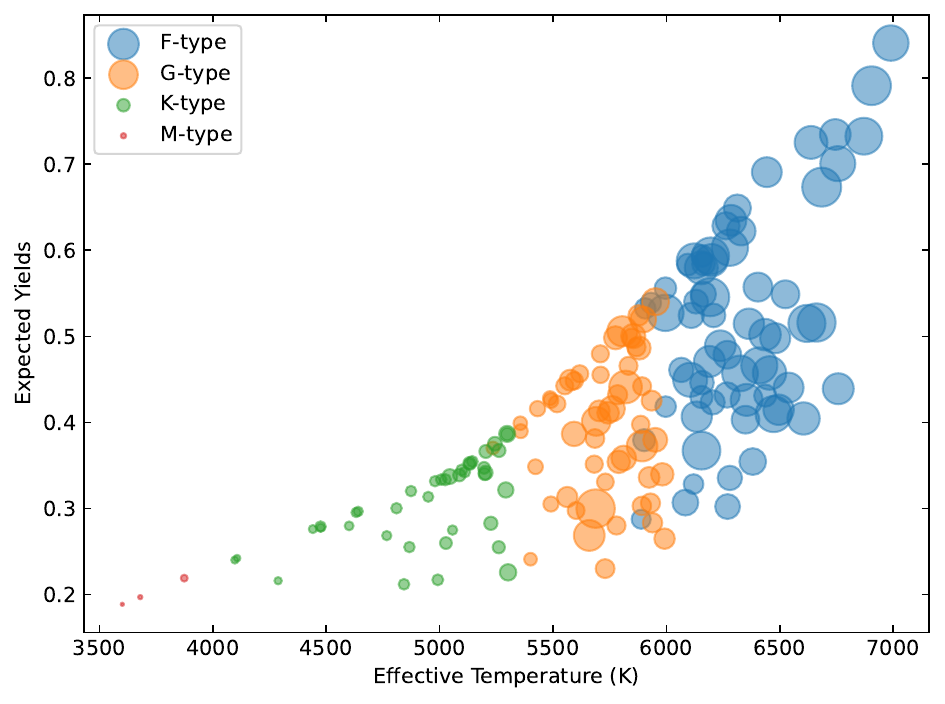}{0.45\textwidth}{(b)}
   }
   \caption{\textit{Left panel}: The histogram of expected number of detected planets versus planetary semi-major axis based on the low frequency and the optimistic habitable zone model. The orange-crosshatched and blue bars represent expected yields by direct imaging alone and imaging+astrometry, respectively. \textit{Right panel}: Planet yields by synergy observations for terrestrial planets in conservative habitable zone as a function of stellar effective temperature. The blue, orange, green and red points represent F, G, K and M stars. The size of each point indicates the position of the habitable zone for the corresponding star.}
   \label{fig:yields}
\end{figure*}

Our results underscore the transformative potential of combining CHES's orbital retrieval capabilities with HWO's imaging power. Figure \ref{fig:yields}(b) presents a histogram of detected planets versus the semi-major axis, based on the low frequency and the optimistic habitable zone model. The yield within each semi-major axis bin increases by approximately one when synergy observations are incorporated. We find the orbital radii of detected planets are primarily concentrated around 1 au, which is also where the enhancements from astrometry are most significant.

{We find that the predicted yields exhibit an increasing trend with stellar effective temperature ($T_\mathrm{eff}$), as shown in the right panel in Figure \ref{fig:yields}(b), which presents the expected yields of terrestrial planets within the conservative habitable zone based on the low occurrence rate model. The rise in planet occurrence rates with increasing $T_\mathrm{eff}$ contributes to this trend. Additionally, the IWA and high precision of HWO enhance the detection completeness, further boosting yields. Similar trends are observed across different occurrence rate assumptions and habitable zone definitions.}

{Based on our results, we assess the priority of each target with astrometry through the following factors \citep{Mamajek2024, Tan2024}}. Firstly, we consider the completeness $C_\mathrm{det}$, efficiency $f_\mathrm{DI+AST}$, and yields $N_p$ based on scenario with astrometry, as higher values of these parameters correspond to greater scientific returns with lower observational time requirements. Thus the priority is calculated based on  $\mathrm{rankindex} = C_\mathrm{det} \times f_\mathrm{DI+AST} \times N_p$. We also adopt the classification  by \cite{Mamajek2024}, for targets with existed disk or a companion star, we multiply the $\mathrm{rankindex}$ by a factor of 0.5 accordingly. For targets with known planets or candidates, we multiply the $\mathrm{rankindex}$ by a factor of 2. The final priority is obtained by normalizing this value and is presented in Table \ref{tab:yield_all}.

\section{Discussion and Conclusion}
\label{sect:dission}

The planning missions, HWO and CHES, are dedicated to discovering terrestrial planets orbiting the nearby stars. With advancements in technology, the detection of habitable terrestrial planets in the optical band appears increasingly feasible in the coming decade. In this work, we explore the synergy between CHES and HWO, demonstrating how their complementary strengths enhance the search for habitable exoplanets.

For all targets of HWO, we inject simulated planets signals and attempt to recover them through imaging alone or combined observations. By leveraging CHES's micro-arcsecond astrometric precision, it becomes possible to predict the optimal observation window for HWO, as illustrated in Figure \ref{fig:predict}. {In this work, we retrieve orbital elements for single-planet systems. However, retrievals for dual-planet systems are also feasible, as demonstrated in our previous studies, where the orbital elements were accurately recovered \citep{Jin2022, Jin2024}. The contrast predictions and observational optimization methods developed here are equally applicable to dual-planet systems, provided that their orbital elements have been resolved. In future work, we will extend our analysis to address orbital retrievals and coordinated observations in more realistic multi-planet systems.} This approach significantly enhances HWO's detection efficiency and planet yields, as summarized in Section \ref{sect:results}. This combined methodology addresses the inherent limitations of each technique: the sensitivity of coronagraph to phase-dependent planetary brightness and astrometry's inability to directly characterize atmospheres.

By leveraging CHES's ability to constrain orbital elements and provide critical data on understanding planetary system architectures, HWO's observational resources can be optimally allocated. The derived orbital parameters, including eccentricity, inclination, and semi-major axis, are essential for assessing system stability, potential habitability, and prospects for biosignature detection \citep{Borges2024}. {This synergy reduces the required number of observations while slightly enhancing the completeness of detecting Earth-like planets.} The findings underscore the importance of multi-method approaches in advancing exoplanetary science. Beyond planet detection, the combined capabilities of CHES and HWO enable detailed studies of planetary atmospheres, compositions, and climates. An expanded sample of Earth-like planets will enhance our understanding of planet formation, evolution, and the conditions necessary for habitability. Moreover, by prioritizing targets identified by CHES, HWO can focus on high-value systems, thereby maximizing the scientific return of the mission.

Despite these promising results, several uncertainties remain that could influence detection yields and conclusions. The final design of HWO's instruments, such as coronagraph performance, throughput, and noise characteristics, will directly impact detection capabilities \citep{Mennesson2024, Steiger2024}. {Since the final target list has not yet been determined, we do not provide an observation schedule. However, similar to the simulations by \citet{Morgan2021}, the schedule can be optimized using astrometry or radial velocity. Here our findings may serve as a valuable reference for selecting priority targets in future studies.} This study considers only single-planet or two-planet systems. However, our methods can be extended to systems with more planets, though this would require additional computational effort. Besides, the final yields are heavily influenced by planet occurrence rates \citep{Stark2024}, which vary widely across different studies \citep{Garrett2018, Zink2019, Pascucci2019, Bryson2021}, Post-processing techniques are not considered in the present study, as listed in \citet{Guyon2006, Lewis2023}, the detection efficiency for planets near the IWA can be reduced due to challenges in distinguishing their signals from the residual point-spread function (PSF). Additionally, the characteristics of target stars (e.g., magnitude, distance and stellar activity) have not been accounted for in orbital retrieval. CHES is expected to achieve higher accuracy in fitting planetary parameters for stars that are nearer and less active, as investigated in our previous study \citep{Bao2024}.

In this work, we adopt a median value of $N_\mathrm{ez}=3$ from the HOSTS survey \citep{Ertel2020}, while improved constraints on exozodiacal dust levels can be considered to refine noise models. \citet{Stark2024} adopted a fitted exozodiacal distribution, and there are three stars with known $N_\mathrm{ez}$, estimated based on detected existed dust: $297\pm56$ around HD 22049, $148\pm28$ around HD 126660 A, and $588\pm121$ around HD 157214 from \citet{Ertel2020}, much higher than the median value. Results of \citet{Salvador2024} show higher $N_\mathrm{ez}$ will make planet yields decrease. Additionally, companion stars in binary systems and circumstellar debris disks can complicate observations \citep{Mamajek2024, Harada2024}. Further observations by CHES could help obtain precise planetary orbits, improving the characterization of these challenging systems.

This study focuses on the optical band around  500~nm. However, HWO's planned multi-band instruments aim to characterize the compositions of planetary atmospheres across a range of wavelengths. Observations optimized for specific bandpasses could further enhance yields \citep{Stark2024b}. Additionally, the yields in the NUV and NIR bands may be lower than the visible band due to instrumental limitations \citep{Morgan2024}.

In conclusion, our work emphasizes the powerful synergy between astrometry and imaging in detecting terrestrial planets in habitable zones. By combining CHES's precise orbital constraints with HWO's imaging capabilities, we can significantly enhance detection efficiency and yields, paving the way for ground-breaking discoveries in exoplanetary science. These results underscore the importance of integrating complementary observational strategies to tackle the complex challenges of identifying and characterizing Earth-like worlds.

In addition to planetary detection, CHES and HWO have the potential to carry out an extensive survey and provide detailed characterization of nearby planetary systems. By expanding the sample of Earth-like planets, these missions will provide valuable insights into the processes of planetary system formation and evolution.

\section*{acknowledgements}
We thank the anonymous reviewer for their insightful comments and suggestions that improved the quality of the manuscript. This work is financially supported by the National Natural Science Foundation of China (grant Nos. 12033010 and 11773081), the Strategic Priority Research Program on Space Science of the Chinese Academy of Sciences (grant No. XDA 15020800), and the Foundation of Minor Planets of the Purple Mountain Observatory.

\software{astropy \citep{2013Astropy},
PyMsOfa \citep{Ji2023},
Scipy \citep{Virtanen2020},
EXOSIMs \citep{Delacroix2016},
Nii-C \citep{Jin2024},
synphot \citep{synphot2018}
}
\hfill

\appendix


\begin{table*}[h!]
   \begin{center}
   \caption{Overall number of detected terrestrial planets} \label{tab:yield}
    \begin{tabular}{ccccc}
   \hline\noalign{\smallskip}
   \hline\noalign{\smallskip}
     & \thead{Imaging \\low occurrence}  &  \thead{Imaging+Astrometry \\low occurrence} & \thead{Imaging \\high occurrence}  & \thead{Imaging+Astrometry \\high occurrence}   \\
   \hline
    conservative HZ & $37.11^{+6.36}_{-6.44}$ & $41.78^{+7.16}_{-7.25}$ & $60.60^{+14.14}_{-13.32}$ & $68.51^{+15.99}_{-15.07}$ \\
   optimistic HZ  & $46.80^{+8.02}_{-8.58}$ & $54.33^{+9.30}_{-9.95}$ & $67.72^{+21.77}_{-19.69}$ & $78.42^{+25.22}_{-22.82}$  \\
   Earth-like  & $16.04^{+3.17}_{-2.85}$ & $17.82^{+3.43}_{-3.09}$ & $25.67^{+7.51}_{-8.10}$ & $28.97^{+8.15}_{-8.80}$ \\
   \hline
   \hline
   \end{tabular}
   \end{center}
\end{table*}

\startlongtable
\centerwidetable
\begin{deluxetable}{lccccccc}
   \tablecaption{Results of terrestrial planets for HWO targets}
\label{tab:yield_all}

\tablehead{
    \multirow{2}{*}{Name} & \multicolumn{3}{c}{Imaging Alone} & \multicolumn{3}{c}{Imaging+Astrometry} & \multirow{2}{*}{\hfill Priority (\%)} \\
    & \colhead{Completeness (\%)} & \colhead{Benefit-to-Cost} & \colhead{Yields} &
    \colhead{Completeness (\%)} & \colhead{Benefit-to-Cost} & \colhead{Yields} &
}
   \startdata
HD 95735&95.42&3.07&0.21  - 0.28&98.40&32.71&0.22 - 0.29&100.00\\
HD 217987&95.37&2.45&0.21 - 0.25&98.46&22.17&0.22 - 0.27&99.61\\
HD 209100&98.97&3.12&0.23 - 0.29&99.68&63.97&0.24 - 0.30&99.23\\
HD 201092&98.52&3.36&0.23 - 0.24&99.26&44.94&0.23 - 0.24&98.85\\
HD 219134&98.71&1.24&0.24 - 0.31&99.66&26.33&0.24 - 0.32&98.46\\
HD 202560&96.05&2.34&0.22 - 0.24&99.01&24.93&0.23 - 0.25&98.08\\
HD 201091&99.03&2.62&0.23 - 0.27&99.28&66.22&0.24 - 0.28&97.69\\
HD 88230&96.62&1.73&0.22 - 0.23&98.87&21.80&0.23 - 0.24&97.31\\
HD 131977&97.46&1.80&0.23 - 0.28&99.35&26.89&0.24 - 0.30&96.92\\
HD 192310&95.05&0.94&0.23 - 0.32&99.44&14.35&0.24 - 0.34&96.54\\
HD 22049&99.85&3.37&0.24 - 0.34&99.92&84.12&0.24 - 0.34&96.15\\
HD 156026&97.04&1.39&0.22 - 0.27&98.72&18.66&0.23 - 0.28&95.77\\
HD 128621&100.00&7.78&0.26 - 0.38&100.00&172.95&0.26 - 0.38&95.38\\
HD 165341 B&97.60&1.60&0.23 - 0.27&98.96&25.20&0.24 - 0.28&95.00\\
HD 10700&99.80&2.85&0.26 - 0.40&99.97&67.47&0.27 - 0.40&94.61\\
HD 185144&99.25&1.87&0.26 - 0.38&99.78&42.02&0.26 - 0.39&94.23\\
HD 4628&95.73&1.35&0.23 - 0.31&99.50&19.19&0.24 - 0.33&93.84\\
HD 10476&97.79&1.18&0.24 - 0.35&99.75&22.86&0.25 - 0.37&93.46\\
HD 26965 A&99.47&1.89&0.25 - 0.35&99.66&40.50&0.25 - 0.35&93.08\\
HD 32147&82.85&1.01&0.18 - 0.24&98.66&11.12&0.23 - 0.30&92.69\\
HD 191408 A&98.50&1.98&0.24 - 0.32&99.33&32.17&0.24 - 0.33&92.31\\
HD 20794&99.40&1.97&0.27 - 0.41&99.93&40.28&0.27 - 0.42&91.92\\
HD 155885&99.31&1.64&0.25 - 0.35&99.65&35.13&0.25 - 0.35&91.54\\
HD 155886&99.18&1.51&0.24 - 0.35&99.54&35.61&0.25 - 0.35&91.15\\
HD 115617&98.36&1.00&0.27 - 0.43&99.85&23.88&0.28 - 0.44&90.77\\
HD 103095&78.35&0.77&0.16 - 0.24&89.26&8.72&0.19 - 0.27&90.38\\
HD 147513&88.81&0.66&0.26 - 0.43&97.76&10.29&0.29 - 0.49&90.00\\
HD 101501&96.31&1.35&0.26 - 0.41&99.37&15.78&0.27 - 0.43&89.61\\
HD 136352&78.08&0.56&0.20 - 0.32&93.36&6.41&0.24 - 0.38&89.23\\
HD 10780&94.69&0.80&0.24 - 0.36&99.14&13.19&0.26 - 0.39&88.84\\
HD 102365&96.34&1.05&0.27 - 0.43&99.67&20.70&0.29 - 0.46&88.46\\
HD 20630&97.94&0.89&0.29 - 0.46&99.79&20.25&0.29 - 0.48&88.07\\
HD 122064&74.93&0.78&0.16 - 0.21&94.06&6.69&0.19 - 0.26&87.69\\
HD 131156 B&77.87&1.12&0.16 - 0.18&96.47&8.78&0.19 - 0.22&87.30\\
HD 3651 A&81.00&0.77&0.19 - 0.27&97.93&9.72&0.24 - 0.34&86.92\\
HD 75732 A&70.29&0.66&0.16 - 0.24&94.96&6.71&0.22 - 0.32&86.53\\
HD 131156 A&99.15&1.38&0.27 - 0.42&99.89&35.24&0.28 - 0.43&86.15\\
HD 166620&73.51&0.63&0.15 - 0.21&91.99&6.33&0.19 - 0.26&85.77\\
HD 128620&99.91&6.95&0.30 - 0.50&99.95&213.09&0.30 - 0.50&85.38\\
HD 69830&72.57&0.66&0.17 - 0.26&95.61&7.21&0.23 - 0.35&85.00\\
HD 156274 A&94.62&1.03&0.23 - 0.34&99.67&17.43&0.25 - 0.37&84.61\\
HD 216803&91.47&0.89&0.20 - 0.25&98.79&12.46&0.23 - 0.28&84.23\\
HD 20807&90.97&0.69&0.27 - 0.45&98.04&11.99&0.30 - 0.50&83.84\\
HD 165341 A&99.68&1.99&0.26 - 0.38&99.86&50.51&0.26 - 0.39&83.46\\
HD 95128&89.37&0.62&0.26 - 0.43&96.44&9.83&0.29 - 0.49&83.07\\
HD 141004&95.55&0.76&0.29 - 0.49&98.77&15.53&0.31 - 0.52&82.69\\
HD 149661&91.99&0.86&0.23 - 0.34&99.18&10.84&0.25 - 0.37&82.30\\
HD 38392&90.69&0.77&0.21 - 0.28&98.51&12.10&0.23  - 0.31&81.92\\
HD 114710&99.02&0.97&0.31 - 0.55&99.70&21.85&0.32 - 0.56&81.53\\
HD 20766&89.81&0.70&0.25 - 0.41&98.12&10.56&0.28 - 0.46&81.15\\
HD 10360&94.66&0.92&0.23 - 0.31&99.53&15.78&0.24 - 0.33&80.76\\
HD 50281&73.29&1.05&0.16 - 0.21&95.31&8.54&0.21 - 0.27&80.38\\
HD 190360&70.53&0.52&0.17 - 0.26&89.47&5.16&0.20 - 0.31&79.99\\
HD 109358&98.86&1.04&0.31 - 0.52&99.84&23.96&0.31 - 0.52&79.61\\
HD 190248&99.68&1.41&0.28 - 0.45&99.90&31.19&0.28 - 0.45&79.22\\
HD 10361&92.84&0.96&0.22 - 0.31&99.58&15.82&0.24 - 0.34&78.84\\
HD 115404 A&65.01&0.81&0.13 - 0.17&88.19&5.66&0.16 - 0.21&78.45\\
HD 43834 A&92.42&1.03&0.25 - 0.40&99.44&15.60&0.28 - 0.45&78.07\\
HD 72673&65.69&0.55&0.14 - 0.21&86.42&5.26&0.17 - 0.26&77.69\\
HD 74576&58.23&0.85&0.11 - 0.16&86.49&5.21&0.16 - 0.22&77.30\\
HD 37394&72.24&0.52&0.16 - 0.23&91.80&5.96&0.20 - 0.28&76.92\\
HD 82885 A&87.29&0.92&0.23 - 0.36&98.81&11.86&0.27 - 0.42&76.53\\
HD 17925&88.22&0.64&0.20 - 0.30&97.66&9.32&0.24 - 0.34&76.15\\
HD 203608&98.15&1.83&0.32 - 0.57&99.72&26.17&0.33 - 0.58&75.76\\
HD 17051&72.28&0.50&0.22 - 0.37&88.61&5.20&0.26 - 0.45&75.38\\
HD 160691&80.15&0.57&0.22 - 0.36&93.05&6.52&0.26 - 0.42&74.99\\
HD 140901 A&63.40&0.43&0.15 - 0.24&85.47&4.22&0.19 - 0.30&74.61\\
HD 14412&65.89&0.62&0.15 - 0.22&84.78&4.86&0.16 - 0.24&74.22\\
HD 4614 A&99.54&2.02&0.31 - 0.53&99.94&47.17&0.31 - 0.53&73.84\\
HD 34411&92.36&0.65&0.28 - 0.46&98.24&12.35&0.30 - 0.50&73.45\\
HD 30495&87.07&0.63&0.25 - 0.41&97.18&8.93&0.28 - 0.47&73.07\\
HD 158633&54.60&0.66&0.12 - 0.18&82.55&4.50&0.15 - 0.23&72.69\\
HD 35296&83.29&0.61&0.27 - 0.47&96.28&9.22&0.31 - 0.54&72.30\\
HD 146233&83.29&0.70&0.23 - 0.37&95.87&7.55&0.26 - 0.43&71.91\\
HD 19373&97.38&1.23&0.31 - 0.52&99.04&21.22&0.31 - 0.54&71.53\\
HD 30652&99.02&1.78&0.36 - 0.69&99.06&42.55&0.36 - 0.69&71.15\\
HD 1581&99.12&1.06&0.31 - 0.53&99.85&24.20&0.31 - 0.54&70.76\\
HD 100623 A&91.20&1.04&0.22 - 0.31&98.98&10.98&0.24 - 0.35&70.38\\
HD 10647&67.86&0.47&0.20 - 0.36&87.85&4.72&0.25 - 0.43&69.99\\
HD 55575&61.59&0.57&0.16 - 0.27&86.30&4.59&0.23 - 0.38&69.61\\
HD 182572&81.49&0.57&0.21 - 0.33&94.29&7.14&0.25 - 0.39&69.22\\
HD 157214&82.67&0.57&0.22 - 0.35&94.84&7.81&0.26 - 0.41&68.84\\
HD 84117&85.05&0.60&0.27 - 0.47&95.34&9.00&0.31 - 0.55&68.45\\
HD 207129&75.00&0.49&0.21 - 0.35&91.89&6.01&0.25 - 0.43&68.07\\
HD 193664&57.15&0.50&0.14 - 0.24&80.69&3.71&0.18 - 0.31&67.68\\
HD 189567&50.50&0.44&0.12 - 0.19&75.71&3.19&0.14 - 0.23&67.30\\
HD 33262 A&96.01&0.79&0.31 - 0.56&99.11&15.62&0.33 - 0.59&66.91\\
HD 142860&97.09&0.76&0.34 - 0.61&97.96&19.20&0.35 - 0.63&66.53\\
HD 72905&71.55&0.69&0.20 - 0.33&94.77&6.61&0.26 - 0.44&66.14\\
HD 143761&67.03&0.51&0.17 - 0.28&87.79&4.92&0.22 - 0.36&65.76\\
HD 33564&62.18&0.45&0.20 - 0.35&79.93&3.91&0.24 - 0.43&65.38\\
HD 160915&82.02&0.39&0.28 - 0.51&89.44&5.41&0.30 - 0.56&64.99\\
HD 32923&79.07&0.54&0.21 - 0.34&91.89&6.71&0.25 - 0.40&64.61\\
HD 38393&98.00&1.73&0.34 - 0.63&99.34&31.69&0.35 - 0.65&64.22\\
HD 206860&53.88&0.45&0.13 - 0.22&77.47&3.17&0.17 - 0.28&63.84\\
HD 69897&72.47&0.49&0.23 - 0.40&87.54&5.06&0.27 - 0.48&63.45\\
HD 140538 A&72.86&0.49&0.18 - 0.29&90.15&5.08&0.22 - 0.35&63.07\\
HD 38858&59.36&0.62&0.14 - 0.23&88.38&4.70&0.21 - 0.33&62.68\\
HD 165499&67.81&0.46&0.19 - 0.32&86.36&4.59&0.23 - 0.38&62.30\\
HD 50692&59.57&0.43&0.16 - 0.26&83.55&3.79&0.20 - 0.34&61.91\\
HD 17206&92.78&0.51&0.32 - 0.58&96.70&10.46&0.34 - 0.62&61.53\\
HD 16895 A&96.72&0.85&0.33 - 0.61&99.07&19.54&0.34 - 0.63&61.14\\
HD 166&68.12&0.48&0.16 - 0.25&90.24&5.08&0.20 - 0.31&60.76\\
HD 78366&46.76&0.39&0.12 - 0.20&72.02&2.73&0.16 - 0.26&60.37\\
HD 90839&89.25&0.85&0.29 - 0.51&98.45&12.31&0.33 - 0.58&59.99\\
HD 4391&73.83&0.45&0.20 - 0.33&89.34&4.92&0.24 - 0.40&59.60\\
HD 165185&58.85&0.57&0.14 - 0.23&82.73&4.06&0.18 - 0.30&59.22\\
HD 53705&68.71&0.50&0.18 - 0.29&87.49&4.77&0.22 - 0.35&58.84\\
HD 43386&67.34&0.43&0.22 - 0.40&82.35&3.98&0.27 - 0.50&58.45\\
HD 7570&84.14&0.59&0.26 - 0.46&95.07&8.28&0.30 - 0.52&58.06\\
HD 65907 A&71.11&0.49&0.21 - 0.35&90.21&5.29&0.25 - 0.42&57.68\\
HD 9826 A&90.79&0.81&0.30 - 0.53&96.51&12.85&0.33 - 0.58&57.30\\
HD 48682&76.24&0.53&0.23 - 0.39&91.21&5.79&0.27 - 0.46&56.91\\
HD 4813&79.11&0.55&0.26 - 0.45&93.39&6.73&0.29 - 0.52&56.53\\
HD 219623&57.35&0.28&0.16 - 0.26&70.27&2.22&0.18 - 0.31&56.14\\
HD 2151&69.33&1.95&0.30 - 0.50&96.92&37.77&0.30 - 0.51&55.76\\
HD 86728 A&74.08&0.69&0.20 - 0.32&93.23&6.82&0.25 - 0.41&55.37\\
HD 197692&91.60&0.60&0.35 - 0.67&96.22&10.77&0.37 - 0.73&54.99\\
HD 110897&55.92&0.55&0.14 - 0.23&80.64&3.59&0.17 - 0.29&54.60\\
HD 142373&82.44&0.58&0.24 - 0.39&92.64&7.41&0.27 - 0.44&54.22\\
HD 739&56.00&0.43&0.17 - 0.32&77.04&3.14&0.23 - 0.41&53.83\\
HD 5015&73.95&0.50&0.22 - 0.38&86.51&5.40&0.26 - 0.45&53.45\\
HD 76151&56.65&0.44&0.14 - 0.22&80.13&3.49&0.17 - 0.28&53.06\\
HD 46588 A&67.08&0.48&0.20 - 0.35&86.03&4.55&0.24 - 0.42&52.68\\
HD 39091&63.26&0.48&0.17 - 0.28&83.51&3.92&0.20 - 0.34&52.29\\
HD 25457&69.57&0.35&0.22 - 0.38&82.44&3.73&0.24 - 0.43&51.91\\
HD 43042&59.37&0.43&0.19 - 0.35&77.60&3.24&0.24 - 0.44&51.52\\
HD 134083&70.50&0.46&0.23 - 0.42&83.56&4.45&0.27 - 0.50&51.14\\
HD 25998&52.62&0.42&0.15 - 0.27&73.53&2.79&0.20 - 0.35&50.75\\
HD 114613&61.36&0.40&0.16 - 0.24&73.61&2.77&0.19 - 0.30&50.37\\
HD 222368&92.74&0.66&0.31 - 0.55&96.31&12.69&0.33 - 0.59&49.99\\
HD 210302&76.57&0.42&0.25 - 0.45&88.50&5.45&0.28 - 0.51&49.60\\
HD 22484&92.05&0.64&0.28 - 0.49&96.36&10.76&0.31 - 0.53&49.22\\
HD 693&69.01&0.54&0.21 - 0.36&85.88&4.50&0.27 - 0.47&48.83\\
HD 84737&66.42&0.50&0.19 - 0.31&85.79&4.60&0.23 - 0.37&48.45\\
HD 125276 A&61.42&0.42&0.16 - 0.28&81.82&3.82&0.19 - 0.33&48.06\\
HD 23249&96.61&0.99&0.24 - 0.33&97.04&22.91&0.24 - 0.34&47.68\\
HD 114837 A&74.39&0.51&0.23 - 0.41&87.95&5.66&0.28 - 0.49&47.29\\
HD 156897 A&84.21&0.54&0.31 - 0.62&91.65&7.28&0.35 - 0.70&46.91\\
HD 58855&60.68&0.43&0.19 - 0.33&78.23&3.45&0.22 - 0.40&46.52\\
HD 64379&75.30&0.47&0.25 - 0.46&86.73&5.00&0.29 - 0.55&46.14\\
HD 128167&88.96&0.60&0.33 - 0.66&95.31&9.40&0.37 - 0.73&45.75\\
HD 187691 A&64.46&0.47&0.20 - 0.34&83.36&4.39&0.24 - 0.41&45.37\\
HD 89449&67.83&0.47&0.21 - 0.38&80.92&4.12&0.26 - 0.47&44.98\\
HD 90589&89.23&0.60&0.35 - 0.72&92.08&9.55&0.39 - 0.79&44.60\\
HD 102870&95.73&0.82&0.32 - 0.57&96.96&21.14&0.33 - 0.59&44.21\\
HD 187013&64.48&0.47&0.21 - 0.37&81.42&4.16&0.25 - 0.46&43.83\\
HD 219482&57.63&0.37&0.16 - 0.28&75.83&3.15&0.19 - 0.34&43.45\\
HD 168151&60.70&0.37&0.19 - 0.35&74.50&2.97&0.22 - 0.41&43.06\\
HD 199260&48.59&0.64&0.13 - 0.23&71.74&2.67&0.17 - 0.30&42.68\\
HD 91324&53.66&0.54&0.15 - 0.25&74.77&3.15&0.21 - 0.37&42.29\\
HD 109085&82.22&0.49&0.32 - 0.64&89.98&6.38&0.36 - 0.73&41.91\\
HD 215648 A&83.09&0.55&0.28 - 0.49&91.28&7.67&0.31 - 0.55&41.52\\
HD 23754&83.72&0.56&0.30 - 0.59&90.99&7.26&0.34 - 0.67&41.14\\
HD 78154 A&65.31&0.49&0.21 - 0.38&81.53&4.29&0.26 - 0.46&40.75\\
HD 105452 A&92.09&0.61&0.38 - 0.78&94.28&10.52&0.40 - 0.84&40.37\\
HD 213845 A&54.71&0.44&0.17 - 0.32&73.58&2.95&0.22 - 0.40&39.98\\
HD 7788 A&61.45&0.38&0.20 - 0.35&75.10&2.93&0.24 - 0.43&39.60\\
HD 160032&66.84&0.44&0.22 - 0.42&79.57&3.73&0.27 - 0.52&39.21\\
HD 212330 A&59.08&0.47&0.14 - 0.22&76.44&3.31&0.17 - 0.27&38.83\\
HD 22001 A&65.83&0.43&0.22 - 0.42&77.20&3.58&0.27 - 0.52&38.44\\
HD 20010 A&94.90&0.65&0.31 - 0.56&96.70&13.01&0.33 - 0.59&38.06\\
HD 90089 A&55.35&0.41&0.18 - 0.35&72.41&2.76&0.23 - 0.44&37.67\\
HD 126660 A&91.45&0.63&0.31 - 0.56&94.83&11.13&0.33 - 0.60&37.29\\
   \enddata
   \tablecomments{The yields are estimated number of planets located in conservative habitable zone, the lower (upper) bounds of yields are based on low (high) occurrence rate model, respectively.}
\end{deluxetable}

\bibliography{ms}
\bibliographystyle{aasjournal}

\listofchanges

\end{document}